\documentclass[english,aps, prl,twocolumn]{revtex4}
\usepackage[T1]{fontenc}
\usepackage[latin9]{inputenc}
\usepackage{babel}
\usepackage{amsmath}
\usepackage{amssymb}
\usepackage{graphicx}
\usepackage{esint}
\usepackage[unicode=true,pdfusetitle,
 bookmarks=true,bookmarksnumbered=false,bookmarksopen=false,
 breaklinks=false,pdfborder={0 0 1},backref=false,colorlinks=false]
 {hyperref}
\usepackage{breakurl}

\makeatletter

\providecommand{\tabularnewline}{\\}

\@ifundefined{textcolor}{}
{%
 \definecolor{BLACK}{gray}{0}
 \definecolor{WHITE}{gray}{1}
 \definecolor{RED}{rgb}{1,0,0}
 \definecolor{GREEN}{rgb}{0,1,0}
 \definecolor{BLUE}{rgb}{0,0,1}
 \definecolor{CYAN}{cmyk}{1,0,0,0}
 \definecolor{MAGENTA}{cmyk}{0,1,0,0}
 \definecolor{YELLOW}{cmyk}{0,0,1,0}
}

\makeatother

\begin{document}

\title{Theory of substrate-directed heat dissipation for single-layer graphene
and other two-dimensional crystals}

\author{Zhun-Yong Ong}

\email{ongzy@ihpc.a-star.edu.sg}

\affiliation{Institute of High Performance Computing, A{*}STAR, Singapore 138632,
Singapore}

\author{Yongqing Cai}

\affiliation{Institute of High Performance Computing, A{*}STAR, Singapore 138632,
Singapore}

\author{Gang Zhang}

\affiliation{Institute of High Performance Computing, A{*}STAR, Singapore 138632,
Singapore}
\begin{abstract}
We present a theory of the phononic thermal (Kapitza) resistance at
the interface between graphene or another single-layer two-dimensional
(2D) crystal (e.g. MoS$_{2}$) and a flat substrate, based on a modified
version of the cross-plane heat transfer model by Persson, Volokitin
and Ueba {[}J. Phys.: Condens. Matter 23, 045009 (2011){]}. We show
how intrinsic flexural phonon damping is necessary for obtaining a
finite Kapitza resistance and also generalize the theory to encased
single-layer 2D crystals with a superstrate. We illustrate our model
by computing the thermal boundary conductance (TBC) for bare and SiO$_{2}$-encased
single-layer graphene and MoS$_{2}$ on a SiO$_{2}$ substrate, using
input parameters from first-principles calculation. The estimated
room temperature TBC for bare (encased) graphene and MoS$_{2}$ on
SiO$_{2}$ are $34.6$ ($105$) and $3.10$ ($5.07$) MWK$^{-1}$m$^{-2}$,
respectively. The theory predicts the existence of a phonon frequency
crossover point, below which the low-frequency flexural phonons in
the bare 2D crystal do not dissipate energy efficiently to the substrate.
We explain within the framework of our theory how the encasement of
graphene with a top SiO$_{2}$ layer introduces new low-frequency
transmission channels which significantly reduce the graphene-substrate
Kapitza resistance. We emphasize that the distinction between bare
and encased 2D crystals must be made in the analysis of cross-plane
heat dissipation to the substrate. 
\end{abstract}
\maketitle

\section{Introduction}

Single-layer two-dimensional (2D) crystals such as graphene and molybdenum
disulphide (MoS$_{2}$) have drawn much interest on account of their
potential for applications in nanoelectronic, optoelectronic and thermoelectric
technology~\cite{Schwierz:NNano10_Graphene,Wang:NNano12_Electronics,Ferrari:Nanoscale15_Science}.
However, the integration of such materials into realistic device configurations
requires them to be in physical contact with other commonly used CMOS
(complementary metal-oxide-semiconductor) technology-compatible materials
such as silicon dioxide (SiO$_{2}$). Although the subnanometer thickness
of such 2D crystals offers considerable advantage in electrostatic
channel scaling in nanoelectronics~\cite{Schwierz:NNano10_Graphene,Liu:ACSNano12_Channel,Lembke:ACR15_Single},
the limited bulk volume is associated with high power densities and
substantial waste Joule heat at high fields~\cite{Meric:NNano08_Current,Serov:JAP14_Theoretical},
which, if not properly dissipated, increases operating temperatures
and has a deleterious effect on device performance and lifetime~\cite{Pop:NResearch10_Energy}.
The waste heat can diffuse laterally within the 2D crystal and/or
directly into the substrate on which the 2D crystal is supported~\cite{Bae:NL10_Imaging}.
In larger devices, heat dissipation to the substrate is the primary
mechanism via which this waste heat is removed. The rate of heat dissipation
depends on the intrinsic voluminal thermal resistance of the substrate
material and the thermal (Kapitza) resistance of the interface between
the 2D crystal and the substrate. Therefore, understanding the physical
mechanism by which the 2D crystal thermally couples to the substrate
is important for the technological development of 2D materials like
graphene and semiconducting transition metal dichalcogenides. However,
we lack a useful theoretical model of heat transfer between 2D crystals
and the three-dimensional (3D) substrate because of their mismatch
in dimensionality. In addition, there are several heat transfer processes
(near-field radiative~\cite{Volokitin:PRB11_Near,Peng:APL15_Thermal},
electron-phonon~\cite{Ong:PRB13_Signature} and phononic~\cite{Persson:JPCM11_Phononic})
between the 2D crystal and the substrate which have been modeled analytically
and numerically. The close agreement between experiments~\cite{Chen:APL09_Chen,Mak:APL10_Measurement}
and molecular dynamics (MD) simulations of the thermal boundary conductance
between graphene and different substrates suggests that heat is dissipated
primarily through phonons (lattice vibrations)~\cite{Ong:PRB11_Effect}. 

In spite of the considerable insight into the interfacial heat transfer
process obtained from MD simulations, we still do not have a direct
physical description of the phononic processes mediating heat dissipation
from a single-layer 2D crystal to the substrate. For example, it is
unclear what role the low-frequency phonons play in cross-plane heat
dissipation. Furthermore, MD simulations are inherently classical
in nature and cannot be related unambiguously to analysis involving
quantum-mechanical phononic processes. Existing theories such as the
acoustic mismatch model (AMM) and the diffuse mismatch model (DMM)
employ analogies to acoustic scattering, specular or diffuse, to describe
phonon transmission at the interface~\cite{Swartz:RMP89_Thermal}.
However, the AMM and the DMM fundamentally assume that bulk incident
phonons are transmitted across or reflected by the interface like
acoustic or electromagnetic waves, a scenario that is not compatible
with the geometrical configuration of supported single or few-layer
2D crystals where there is no extended volume in the direction normal
to the interface. Therefore, it is necessary to formulate a fresh
theory that takes into account the mismatch in dimensionality between
the 2D crystal and its 3D substrate, and explicitly describes the
phononic and vibrational character of such structures. 

In our theory, we do not assume that the 2D crystal has any extended
volume in the out-of-plane direction like in conventional mismatch
models. Indeed, this approach offers us the unique advantage of directly
linking the flexural nature of the 2D crystal to phonon transmission
to the substrate. Moreover, this allows us to incorporate the effects
of a superstrate, such as a top gate oxide layer, on heat dissipation
to the \emph{substrate}, and enables us to understand the difference
in heat dissipation to the substrate by a bare 2D crystal and that
by its encased counterpart. Our theory predicts that encased 2D crystals
have a significantly higher thermal boundary conductance (or lower
Kapitza resistance), consistent with empirical trends observed across
different experiments. In spite of the simplicity of the model, its
numerical predictions of the TBC are in very good agreement with published
experimental data for graphene and MoS$_{2}$, using numerical parameters
obtained from first-principles calculations and published elasticity
parameters.

The organization of our paper is as follows. We begin with the derivation
of our theory of flexural phonon-mediated interfacial heat transfer
for bare and encased graphene, and show how our theory is obtained
by modifying the model by Persson, Volokitin and Ueba~\cite{Persson:JPCM11_Phononic}.
We then discuss how the damping function for flexural phonons, a key
element of our theory, can be approximated and how the interfacial
spring constant can be estimated in density functional theory (DFT)
calculations. We apply the theory to estimate the TBC values for bare
and SiO$_{2}$-encased graphene and MoS$_{2}$, and compare them with
published experimental~\cite{Freitag:NL09_Energy,Chen:APL09_Chen,Mak:APL10_Measurement,Taube:AMI15_Temperature}
and simulation~\cite{Ni:APL13_Few} data. Excellent agreement is
obtained between our predicted TBC values and the various experimental
and simulation data. We also discuss the physics underlying the higher
TBC of SiO$_{2}$-encased graphene and MoS$_{2}$, and interpret it
in terms of the additional low-frequency transmission channels due
to coupling to the superstrate. Finally, we give an overall picture
of the heat dissipation pathways in supported 2D crystals and discuss
how the TBC may change when the different components are modified.

\section{Theory of heat transfer\label{Sec:HeatTransferTheory}}

\begin{figure}
\includegraphics[width=7.5cm]{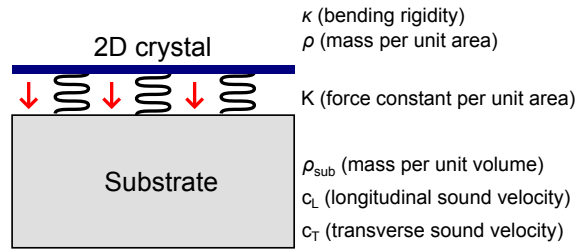}

\caption{Schematic representation of the bare single-layer 2D crystal (graphene
or MoS$_{2}$) and the substrate. The springs represent the weak van
der Waals interaction at the interface. The simulation parameters
are also displayed.}
\label{Fig:ModelSchematics}
\end{figure}

\subsection{Bare single-layer 2D crystal and solid substrate}

The point of departure in the formulation of our theory is Ref.~\cite{Persson:JPCM11_Phononic}
by Persson, Volokitin and Ueba. We begin with a self-contained introduction
to the essential elements of Ref.~\cite{Persson:JPCM11_Phononic},
following closely the treatment in Ref.~\cite{Persson:JPCM11_Phononic}.
Suppose we have a single-layer 2D crystal supported by a flat substrate
as shown in Fig.~\ref{Fig:ModelSchematics}, with the $z$-axis perpendicular
to the substrate surface. The normal stress acting on the 2D crystal
at position $\mathbf{r}=(x,y)$ and time $t$ is 
\[
\sigma_{\text{int}}(\mathbf{r},t)=-K[u_{\text{2D}}(\mathbf{r},t)-u_{\text{sub}}(\mathbf{r},t)]
\]
where $u_{\text{2D}}$ and $u_{\text{sub}}$ are respectively the
out-of-plane displacement of the 2D crystal and the substrate surface,
and $K$ is the spring constant per unit area characterizing the interaction,
typically van der Waals, at the interface. If we take the Fourier
transform of $u(\mathbf{r},t)$, \emph{i.e.}, $u(\mathbf{q},\omega)=(2\pi)^{-3}\int d^{2}r\int dt\ u(\mathbf{r},t)e^{i(\mathbf{q}\cdot\mathbf{r}-\omega t)}$
where $\mathbf{q}$ and $\omega$ are respectively the crystal momentum
and frequency, then we have
\begin{equation}
\sigma_{\text{int}}(\mathbf{q},\omega)=-K[u_{\text{2D}}(\mathbf{q},\omega)-u_{\text{sub}}(\mathbf{q},\omega)]\ .\label{Eq:InterfaceForce}
\end{equation}
The equation of motion for the 2D crystal is 
\begin{align}
 & \rho\frac{\partial^{2}u_{\text{2D}}(\mathbf{r},t)}{\partial t^{2}}+\rho\gamma\frac{\partial u_{\text{2D}}(\mathbf{r},t)}{\partial t}+\kappa\nabla^{2}\nabla^{2}u_{\text{2D}}(\mathbf{r},t)\nonumber \\
 & =\sigma_{\text{int}}(\mathbf{r},t)+\sigma_{\text{f}}(\mathbf{r},t)\label{Eq:EquationMotion_1}
\end{align}
where $\rho$ and $\kappa$ are the mass density per unit area and
the bending stiffness of the uncoupled 2D crystal, respectively and
$\gamma$ is the damping coefficient representing the \emph{intrinsic}
damping of the flexural motion. On the RHS of Eq.~(\ref{Eq:EquationMotion_1}),
$\sigma_{\text{f}}(\mathbf{r},t)$ represents the stochastic force
from thermal fluctuation within the 2D crystal, in addition to the
interface force from the substrate. The Fourier transform of Eq.~(\ref{Eq:EquationMotion_1})
yields the algebraic expression 
\begin{align}
 & -(\rho\omega^{2}+i\rho\gamma\omega-\kappa q^{4})u_{\text{2D}}(\mathbf{q},\omega)\nonumber \\
 & =\sigma_{\text{int}}(\mathbf{q},\omega)+\sigma_{\text{f}}(\mathbf{q},\omega)\label{Eq:EquationMotion_1_operator}
\end{align}
or
\begin{equation}
u_{\text{2D}}(\mathbf{q},\omega)=u_{\text{f}}(\mathbf{q},\omega)-D_{\text{2D}}(\mathbf{q},\omega)\sigma_{\text{int}}(\mathbf{q},\omega)\ ,\label{Eq:TransferFunction_1}
\end{equation}
where 
\begin{equation}
D_{\text{2D}}(\mathbf{q},\omega)=\lim_{\eta\rightarrow0^{+}}(\rho\omega^{2}+i\rho\gamma\omega-\kappa q^{4}+i\eta)^{-1}\label{Eq:GreensFunction_1}
\end{equation}
is the \emph{retarded} Green's function~\cite{Amorim:PRB13_Flexural}
for the flexural motion of the 2D crystal at the interface, and $u_{\text{f}}(\mathbf{q},\omega)=-D_{\text{2D}}(\mathbf{q},\omega)\sigma_{\text{f}}(\mathbf{q},\omega)$
is the stochastic component of the flexural motion due to thermal
fluctuation. Equation~(\ref{Eq:TransferFunction_1}) describes the
flexural response of the 2D crystal to a periodic harmonic stress
exerted at the interface and the stochastic force. We modify Eq.~(\ref{Eq:GreensFunction_1})
to take into account frequency-dependent damping by setting $\gamma$
as a function of $\omega$, \emph{i.e.} $\gamma=\gamma(\omega)$.
The key difference between our model and the one in Ref.~\cite{Persson:JPCM11_Phononic}
is our inclusion of the damping process represented by the second
term on the right hand side of Eq.~(\ref{Eq:GreensFunction_1}).
As we will show later, it allows us to avoid the weak coupling approximation
in Ref.~\cite{Persson:JPCM11_Phononic} which can result in numerically
inaccurate estimates of the Kapitza resistance. This damping term
represents phenomenologically the coupling and exchange of energy
between the flexural phonons with the other \emph{intrinsic} degrees
of freedom (e.g. electrons and in-plane acoustic phonons) in the 2D
crystal. Physically, this means that a time-dependent applied force
at the interface results in the excitation of flexural modes but the
energy of these flexural phonons eventually dissipates to and is adsorbed
by the other intrinsic degrees of freedom to which the flexural phonons
are coupled. In the language of many-body physics, the damping term
corresponds to the self-energy from the interaction of the flexural
phonons with other intrinsic degrees of freedom. 

The response of the substrate surface to the interfacial stress is
similarly expressed as
\begin{equation}
u_{\text{sub}}(\mathbf{q},\omega)=D_{\text{\text{sub}}}(\mathbf{q},\omega)\sigma_{\text{int}}(\mathbf{q},\omega)\label{Eq:SubstrateResponse}
\end{equation}
where $D_{\text{sub}}(\mathbf{q},\omega)$ is the retarded Green's
function for the free surface displacement of the isotropic solid
substrate. Within the elastic continuum model, for an elastic solid
with isotropic elastic properties, we have~\cite{Persson:JCP01_Theory,Persson:JPCM11_Phononic}
\begin{equation}
D_{\text{\text{sub}}}(\mathbf{q},\omega)=\frac{i}{\rho_{\text{sub}}c_{T}^{2}}\frac{p_{L}(q,\omega)}{S(q,\omega)}\left(\frac{\omega}{c_{T}}\right)^{2}\label{Eq:SubstrateGreensFunction}
\end{equation}
where \begin{subequations}
\begin{equation}
S(q,\omega)=\left[\left(\frac{\omega}{c_{T}}\right){}^{2}-2q^{2}\right]{}^{2}+4q^{2}p_{T}p_{L}\ ,
\end{equation}
\begin{equation}
p_{L}(q,\omega)=\lim_{\eta\rightarrow0^{+}}\left[\left(\frac{\omega}{c_{L}}\right)^{2}-q^{2}+i\eta\right]^{1/2}\ ,
\end{equation}
\begin{equation}
p_{T}(q,\omega)=\lim_{\eta\rightarrow0^{+}}\left[\left(\frac{\omega}{c_{T}}\right)^{2}-q^{2}+i\eta\right]^{1/2}\ ,
\end{equation}
\end{subequations}and $c_{L}$, $c_{T}$ and $\rho_{\text{sub}}$
are the longitudinal and transverse velocities, and the mass density
per unit volume, respectively. Given the stochastic thermal fluctuation
in the 2D crystal, the resultant motion of the 2D crystal and the
substrate can be obtained by combining Eqs.~(\ref{Eq:TransferFunction_1})
and (\ref{Eq:SubstrateResponse}) to yield: \begin{subequations}
\begin{equation}
u_{\text{sub}}(\mathbf{q},\omega)=\frac{-KD_{\text{sub}}(\mathbf{q},\omega)u_{\text{f}}(\mathbf{q},\omega)}{1-K[D_{\text{2D}}(\mathbf{q},\omega)+D_{\text{sub}}(\mathbf{q},\omega)]}\ ,\label{Eq:SubstrateDisplacement}
\end{equation}
\begin{equation}
u_{\text{2D}}(\mathbf{q},\omega)=\frac{[1-KD_{\text{sub}}(\mathbf{q},\omega)]u_{\text{f}}(\mathbf{q},\omega)}{1-K[D_{\text{2D}}(\mathbf{q},\omega)+D_{\text{sub}}(\mathbf{q},\omega)]}\ .\label{Eq:2DCrystalDisplacement}
\end{equation}
\label{SubEq:Displacements}\end{subequations} 

Given the thermal fluctuation in the flexural motion of the 2D crystal,
this stochastic motion also causes the interfacial stress on the substrate
surface to fluctuate and transmit energy back and forth between the
2D crystal and the substrate. The associated thermal energy transfer
from the 2D crystal to the substrate over the time period $\tau$
is
\[
\Delta E_{\text{2D}\rightarrow\text{sub}}=-\int d^{2}r\int_{0}^{\tau}dt\ \frac{\partial u_{\text{sub}}(\mathbf{r},t)}{\partial t}\sigma_{\text{int}}(\mathbf{r},t)\ ,
\]
where the spatial integration is over the entire plane of the interface.
One can also write
\begin{align}
\Delta E_{\text{2D}\rightarrow\text{sub}}= & \frac{1}{(2\pi)^{3}}\int d^{2}q\int d\omega\nonumber \\
 & \times i\omega u_{\text{sub}}(\mathbf{q},\omega)\sigma_{\text{int}}(-\mathbf{q},-\omega)\ .\label{Eq:EnergyTransfer_1}
\end{align}
Substituting Eqs.~(\ref{Eq:InterfaceForce}) and (\ref{SubEq:Displacements})
in Eq.~(\ref{Eq:EnergyTransfer_1}), we have the expression for the
average thermal energy transfer, \emph{i.e.},
\begin{align}
\langle\Delta E_{\text{2D}\rightarrow\text{sub}}\rangle= & -\frac{1}{(2\pi)^{3}}\int d^{2}q\int d\omega\nonumber \\
 & \times\frac{\omega K^{2}\text{Im}D_{\text{sub}}(\mathbf{q},\omega)\langle|u_{\text{f}}(\mathbf{q},\omega)|^{2}\rangle}{|1-K[D_{\text{sub}}(\mathbf{q},\omega)+D_{\text{2D}}(\mathbf{q},\omega)]|^{2}}\ \label{Eq:AverageEnergyTransfer_1}
\end{align}
where $\langle\ldots\rangle$ denotes the thermal average. We can
write $\langle|u_{\text{f}}(\mathbf{q},\omega)|^{2}\rangle$ as the
Fourier transform of the autocorrelation function of the displacement
$\langle u_{\text{f}}(\mathbf{r},t)u_{\text{f}}(0,0)\rangle$, \emph{i.e.},
\begin{equation}
\langle|u_{\text{f}}(\mathbf{q},\omega)|^{2}\rangle=\frac{A\tau}{(2\pi)^{3}}C_{uu}(\mathbf{q},\omega)\label{Eq:AutocorrelationFunction}
\end{equation}
where 
\[
C_{uu}(\mathbf{q},\omega)=\frac{1}{(2\pi)^{3}}\int d^{2}r\int dt\langle u_{\text{f}}(\mathbf{r},t)u_{\text{f}}(0,0)\rangle e^{i(\mathbf{q}\cdot\mathbf{r}-\omega t)}\ 
\]
and $A$ is the area of the interface. The fluctuation-dissipation
theorem~\cite{Persson:JPCM11_Phononic} implies that 
\begin{equation}
C_{uu}(\mathbf{q},\omega)=-\frac{2\hbar N(\omega,T)}{(2\pi)^{3}}\text{Im}D_{\text{2D}}(\mathbf{q},\omega)\label{Eq:FluctDissip}
\end{equation}
where $N(\omega,T)=[\exp(\hbar\omega/k_{B}T)-1]^{-1}$ is the Bose-Einstein
occupation function for frequency $\omega$ at temperature $T$. The
heat current from the 2D crystal to the substrate due to the thermal
fluctuations in the 2D crystal is $J_{\text{2D}\rightarrow\text{sub}}=\langle\Delta E_{\text{2D}\rightarrow\text{sub}}\rangle/(A\tau)$.
Thus, Eq.~(\ref{Eq:AverageEnergyTransfer_1}) gives us the expression
for the average power dissipated from the 2D crystal to the substrate:
\begin{align*}
J_{\text{2D}\rightarrow\text{sub}}(T)= & \frac{4}{(2\pi)^{3}}\int d^{2}q\int_{0}^{\infty}d\omega\hbar\omega N(\omega,T)\\
 & \times\frac{K^{2}\text{Im}D_{\text{2D}}(\mathbf{q},\omega)\text{Im}D_{\text{sub}}(\mathbf{q},\omega)}{|1-K[D_{\text{2D}}(\mathbf{q},\omega)+D_{\text{sub}}(\mathbf{q},\omega)]|^{2}}\ .
\end{align*}
Using the same procedure, a similar expression can be obtained for
the average power dissipated from the substrate to 2D crystal, $J_{\text{sub}\rightarrow\text{2D}}(T)$.
Therefore, the thermal boundary conductance $G$ for the interface~\cite{Persson:JPCM11_Phononic}
is $G(T)=\lim_{\delta T\rightarrow0}[J_{\text{2D}\rightarrow\text{sub}}(T+\delta T/2)-J_{\text{sub}\rightarrow\text{2D}}(T-\delta T/2)]/\delta T$,
giving us

\begin{align}
G(T)= & \frac{4K^{2}}{(2\pi)^{3}}\int d^{2}q\int_{0}^{\infty}d\omega\hbar\omega\frac{\partial N(\omega,T)}{\partial T}\nonumber \\
 & \times\frac{\text{Im}D_{\text{sub}}(\mathbf{q},\omega)\text{Im}D_{\text{2D}}(\mathbf{q},\omega)}{|1-K[D_{\text{sub}}(\mathbf{q},\omega)+D_{\text{2D}}(\mathbf{q},\omega)]|^{2}}\ .\label{Eq:HeatTransfer_1Sheet}
\end{align}
We can express Eq.~(\ref{Eq:HeatTransfer_1Sheet}) in the more familiar
Landauer form
\begin{equation}
G(T)=\frac{1}{(2\pi)^{3}}\int d^{2}q\int d\omega\hbar\omega\frac{\partial N(\omega,T)}{\partial T}\Xi(q,\omega)\label{Eq:HeatTransfer_Landauer}
\end{equation}
where 
\begin{align}
\Xi(\mathbf{q},\omega)= & \frac{4K^{2}\text{Im}D_{\text{sub}}(\mathbf{q},\omega)\text{Im}D_{\text{2D}}(\mathbf{q},\omega)}{|1-K[D_{\text{sub}}(\mathbf{q},\omega)+D_{\text{2D}}(\mathbf{q},\omega)]|^{2}}\label{Eq:TransmissionFunction}
\end{align}
is the transmission function at $(\mathbf{q},\omega)$. If we assume
that $\text{Im}D_{\text{sub}}(\mathbf{q},\omega)\geq0$ and $\text{Im}D_{\text{2D}}(\mathbf{q},\omega)\geq0$,
we can show in Appendix~\ref{Sec:NumericalBoundsTransmission} that
$0\leq\Xi(\mathbf{q},\omega)\leq1$ which is exactly the property
needed for a transmission coefficient. 

Here, we comment on the weak coupling approximation originally used
in Ref.~\cite{Persson:JPCM11_Phononic} for evaluating Eq.~(\ref{Eq:HeatTransfer_1Sheet}).
It is claimed~\cite{Persson:JPCM11_Phononic} that the denominator
in Eq.~(\ref{Eq:TransmissionFunction}) can be ignored when the coupling
between the two solids is weak (\emph{i.e.}, $K$ is so small that
$|K[D_{\text{sub}}(\mathbf{q},\omega)+D_{\text{2D}}(\mathbf{q},\omega)]|\ll1$)
and that the numerator in Eq.~(\ref{Eq:TransmissionFunction}) is
thus dominated by the pole contribution from $\text{Im}D_{\text{sub}}(\mathbf{q},\omega)\text{Im}D_{\text{2D}}(\mathbf{q},\omega)$
at the $(q,\omega)$-point where the dispersion curves for the substrate
Rayleigh modes and the bending modes of the 2D crystal intersect.
However, this argument does not hold because the denominator $|1-K[D_{\text{sub}}(\mathbf{q},\omega)+D_{\text{2D}}(\mathbf{q},\omega)]|^{2}$
diverges along the poles of $D_{\text{sub}}(\mathbf{q},\omega)$ and
$D_{\text{2D}}(\mathbf{q},\omega)$ and there are actually no singularities
in $\Xi(\mathbf{q},\omega)$ which is strictly less than or equal
to unity (see Appendix~\ref{Sec:NumericalBoundsTransmission} for
the proof). Therefore, the weak-coupling approximation is incorrect
and the denominator in Eqs.~(\ref{Eq:TransmissionFunction}) and
(\ref{Eq:HeatTransfer_1Sheet}) must be retained to avoid spurious
singularities. Our calculations of the integral {[}Eq.~(\ref{Eq:HeatTransfer_1Sheet}){]}
without the weak-coupling approximation indicate that the integral
in Eq.~(\ref{Eq:HeatTransfer_1Sheet}) converges numerically to zero
if there is no damping. This is because $\Xi(\mathbf{q},\omega)$
is only nonzero when along the poles of $D_{\text{sub}}(\mathbf{q},\omega)$
and $D_{\text{2D}}(\mathbf{q},\omega)$ when there is no damping,
i.e. $\gamma(\omega)=0$. Elsewhere, the numerator in the RHS of Eq.~(\ref{Eq:TransmissionFunction})
and hence, $\Xi(\mathbf{q},\omega)$ are zero. Physically, this means
that a dissipative mechanism must be present for net heat transfer
to take place. This invalid approximation also explains why the numerical
estimate~\cite{Persson:JPCM11_Phononic} of the TBC for the graphene/SiO$_{2}$
interface is an order of magnitude too large.

\subsection{Single-layer 2D crystal with superstrate and solid substrate}

In more practical device designs, the perpendicular electrostatic
field is applied through a top gate which consists of a metal insulated
from the channel by a layer of oxide such as SiO$_{2}$ or HfO$_{2}$.
Top-gated graphene and MoS$_{2}$ field-effect transistors are known
to have superior carrier density modulation and higher mobilities~\cite{Fallahazad:APL10_Dielectric,Fallahazad:APL12_Scaling,Ong:PRB12_Charged,Ong:APL13_Top,Radisavljevic:NMat13_Mobility,Ong:PRB13_Mobility}
because of better charge screening. However, the effect of the top
oxide layer, the \emph{superstrate}, on heat dissipation from the
encased 2D crystal to the substrate has not been systematically studied
or even considered. 

\begin{figure}
\includegraphics[width=5cm]{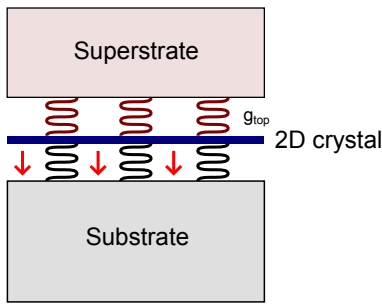}

\caption{Schematic of the single-layer 2D crystal (graphene or MoS$_{2}$)
with the superstrate and the substrate. }
\label{Fig:ModelSchematics_Superstrate}
\end{figure}

Given the theory of heat transfer between the \emph{bare} 2D crystal
and the substrate surface, it is possible incorporate the effects
of a superstrate, as shown in Fig.~\ref{Fig:ModelSchematics_Superstrate},
on the 2D crystal by modifying the equation of motion in Eq.~(\ref{Eq:EquationMotion_1_operator}).
Let $u_{\text{top}}(\mathbf{q},\omega)$ be the out-of-plane displacement
of the bottom surface of the superstrate and $\sigma_{\text{top}}=g_{\text{top}}[u_{\text{2D}}(\mathbf{q},\omega)-u_{\text{top}}(\mathbf{q},\omega)]$,
where $g_{\text{top}}$ is the spring constant per unit area characterizing
the coupling to the superstrate, be the force per unit area exerted
by the superstrate on the 2D crystal. Therefore, the equation of motion
for the 2D crystal is
\begin{align}
 & -(\rho\omega^{2}+i\rho\gamma\omega-\kappa q^{4})u_{\text{2D}}(\mathbf{q},\omega)\nonumber \\
 & =\sigma_{\text{int}}(\mathbf{q},\omega)-\sigma_{\text{top}}(\mathbf{q},\omega)+\sigma_{\text{f}}(\mathbf{q},\omega)\ .\label{Eq:EquationMotion_Covered}
\end{align}
 Like in Eq.~(\ref{Eq:SubstrateResponse}), we define the surface
response of the superstrate as
\begin{equation}
u_{\text{top}}(\mathbf{q},\omega)=D_{\text{\text{top}}}(\mathbf{q},\omega)\sigma_{\text{top}}(\mathbf{q},\omega)\label{Eq:SuperstrateResponse}
\end{equation}
where $D_{\text{top}}(\mathbf{q},\omega)$ is the retarded Green's
function for the free surface displacement of the superstrate. We
combine Eqs.~(\ref{Eq:EquationMotion_Covered}) and (\ref{Eq:SuperstrateResponse})
to obtain the \emph{effective} equation of motion for the 2D crystal
analogous to Eq.~(\ref{Eq:EquationMotion_1_operator}) 
\begin{align}
 & -\left[\rho\omega^{2}+i\rho\gamma\omega-\kappa q^{4}-P(\mathbf{q},\omega)\right]u_{\text{2D}}(\mathbf{q},\omega)\nonumber \\
 & =\sigma_{\text{int}}(\mathbf{q},\omega)+\sigma_{\text{f}}(\mathbf{q},\omega)\label{Eq:EffectiveEquationMotion_Covered}
\end{align}
where $P(\mathbf{q},\omega)=g_{\text{top}}[1-g_{\text{top}}D_{\text{\text{top}}}(\mathbf{q},\omega)]^{-1}$
is the `self-energy' contribution from coupling to the superstrate.
Therefore, we define the effective retarded Green's function for the
flexural motion of the covered 2D crystal as 
\begin{equation}
\overline{D}_{\text{2D}}(\mathbf{q},\omega)=[\rho\omega^{2}+i\rho\gamma\omega-\kappa q^{4}-P(\mathbf{q},\omega)]^{-1}\ .\label{Eq:GreensFunction_Covered}
\end{equation}
In the case of a top oxide layer, $D_{\text{top}}(\mathbf{q},\omega)$
may have the same functional form as Eq.~(\ref{Eq:SubstrateGreensFunction}).
In fact, if the superstrate is of the same material as the substrate,
we can set $D_{\text{top}}(\mathbf{q},\omega)=D_{\text{sub}}(\mathbf{q},\omega)$
and $g_{\text{top}}=K$. For a thin film of adsorbates on the 2D crystal,
we can write $D_{\text{top}}(\mathbf{q},\omega)=1/(\rho_{\text{top}}\omega^{2})$
where $\rho_{\text{top}}$ is mass density of the adsorbates. In our
simulations, we assume that the superstrate material is SiO$_{2}$
like the substrate and the interfacial spring constant per unit area
for the top and bottom interfaces of the 2D crystal are equal. Hence,
like Eq.~(\ref{Eq:HeatTransfer_1Sheet}), the expression for the
thermal boundary conductance is

\begin{align}
G(T)= & \frac{4K^{2}}{(2\pi)^{3}}\int d^{2}q\int_{0}^{\infty}d\omega\hbar\omega\frac{\partial N(\omega,T)}{\partial T}\nonumber \\
 & \times\frac{\text{Im}D_{\text{sub}}(\mathbf{q},\omega)\text{Im}\overline{D}_{\text{2D}}(\mathbf{q},\omega)}{|1-K[D_{\text{sub}}(\mathbf{q},\omega)+\overline{D}_{\text{2D}}(\mathbf{q},\omega)]|^{2}}\ .\label{Eq:HeatTransfer_1Sheet_TopOxide}
\end{align}

\section{Additional details on numerical calculations}

Although the details of our theory of heat transfer are described
in Section~\ref{Sec:HeatTransferTheory}, there are two other important
elements that are needed for the calculation of the thermal boundary
conductance and warrant a more in-depth discussion before the theory
can be made useful for numerical calculations. The first is the function
form of the damping function for the flexural phonons while the second
is the spring constant at the interface. The other material parameters
such as the bending rigidity ($\kappa$) and the 2D crystal mass density
($\rho$) can be obtained from literature.

\subsection{Damping function for flexural phonons}

A key element of our theory is the form of the damping function $\gamma(\omega)$
that determines the rate at which the flexural (ZA) phonon dissipates
energy internally to other intrinsic degrees of freedom and is necessary
to produce a finite Kapitza resistance. First principles calculations
of the room-temperature anharmonic phonon-phonon scattering rate for
long-wavelength flexural phonons in single-layer graphene by Bonini
\emph{et al.}~\cite{Bonini:NL12_Acoustic} suggest that $\gamma(\omega)\propto\omega$,
\emph{i.e.}, the flexural phonon lifetime $\tau(\omega)$ is proportional
to its period. The same linear relationship between the scattering
rate and the frequency of the long-wavelength flexural phonons is
also observed in Ref.~\cite{Lindsay:PRB14_Phonon}. In this work,
we ignore the effects of direct electron-phonon coupling on flexural
phonon scattering since the interaction is a second-order effect~\cite{Viljas:PRB10_Electron,Castro:PRL10_Limits,Mariani:PRL08_Flexural}
and the flexural phonon lifetimes in graphene can be satisfactorily
explained~\cite{Lindsay:PRB10_Flexural,Lindsay:PRB14_Phonon} without
including the effects of flexural phonon scattering with electrons.
Moreover, there is close agreement for the thermal conductivity in
graphene between experimental data and theoretical estimates~\cite{Lindsay:PRB10_Flexural,Lindsay:PRB14_Phonon}
that are calculated using only anharmonic phonon coupling and without
any electron-phonon interactions. Thus, the electron-phonon contribution
to flexural phonon damping is small and can be neglected. Likewise,
the thermal conductivity of MoS$_{2}$ can also be modeled without
needing to consider the effects of electron-phonon scattering on the
acoustic phonon lifetimes~\cite{Li:APL13_Thermal,Wei:APL14_Phonon}.

Given that $\gamma(\omega)=2/\tau(\omega)$ for a damped harmonic
oscillator, we estimate from Ref.~\cite{Bonini:NL12_Acoustic} that
$\omega/\gamma(\omega)\approx115$ in graphene at room temperature
($T_{\text{RT}}=300$ K) although the ratio is slightly lower at $\sim90$
in Ref.~\cite{Lindsay:PRB14_Phonon}. The quality factor in graphene
nanoresonators has also been estimated to be around $\sim100$ and
to scale as $T^{-1}$ in molecular dynamics simulations~\cite{Jiang:Nanoscale14_MoS2}.
Therefore, we propose a phenomenological expression for the temperature-dependent
$\gamma(\omega)$ in graphene:
\begin{equation}
\gamma(\omega,T)=\frac{\omega T}{\alpha T_{\text{RT}}}\label{Eq:DampingFunc}
\end{equation}
where $\alpha$ is equal to the ratio of the phonon lifetime to its
period at room temperature. In our simulations, we set to $\alpha=100$
for single-layer graphene and MoS$_{2}$. In the case of MoS$_{2}$,
there is less certainty over the value of $\alpha$ as there is no
readily available data on the relationship between flexural phonon
lifetime and frequency although molecular dynamics simulations in
Ref.~\cite{Jiang:Nanoscale14_MoS2} show that the quality factor
for MoS$_{2}$ is about three times that for graphene, suggesting
that $\alpha$ can be as high as $\sim300$. We note that in spite
of the high flexural phonon lifetimes in MoS$_{2}$, its thermal conductivity
is much lower than that of graphene because of the low phonon group
velocities in MoS$_{2}$ and the stronger anharmonicity for the in-plane
longitudinal and transverse acoustic phonons~\cite{Li:APL13_Thermal,Wei:APL14_Phonon}. 

The functional form of Eq.~(\ref{Eq:DampingFunc}) also means that
flexural phonon damping increases linearly with temperature~\cite{Klemens:Book58_Thermal}.
We stress that the functional form of Eq.~(\ref{Eq:DampingFunc})
is approximate and that its frequency and temperature dependence can
be in principle more precisely evaluated through DFT calculations
and more sophisticated many-body techniques although this is beyond
the scope of our current work. The damping is also expected to be
higher when the graphene or MoS$_{2}$ has \emph{intrinsic} defects
such as grain boundaries or vacancies that can reduce the flexural
phonon lifetimes through elastic scattering.

\subsection{Density functional theory calculation-based estimate of spring constant
at interface}

Another important parameter needed for our numerical calculation is
$K$, the spring constant per unit area for the interface. The numerical
value of $K$ depends on the 2D crystal, the substrate and the surface
atomistic structure of the substrate. We estimate the numerical values
of $K$ from first principles calculations by computing the energy
change when the 2D crystal is displaced from its equilibrium distance
$d_{0}$ to the substrate surface. The change in energy per unit area
for small displacements $d-d_{0}$ is thus 
\[
\frac{\Delta E}{A}=\frac{1}{2}K(d-d_{0})^{2}\ ,
\]
where $\Delta E$ is the energy change for the interface, $A$ is
the area of the interface and $d$ is the distance between the 2D
crystal and the substrate surface. 

We estimate $K$ for graphene and MoS$_{2}$ on H- and OH-terminated
SiO$_{2}$ from density functional theory (DFT) calculations. The
interlayer spacing-energy ($d-E$) profiles for the various heterostructures
(graphene/SiO$_{2}$ and MoS$_{2}$/SiO$_{2}$) are calculated with
the first-principles method within the framework of density functional
theory by using the software package VASP~\cite{Kresse:PRB96_Iterative}.
The DFT-D2 method is adopted to simulate the van der Waals interactions
across the interface. The Perdew-Burke-Ernzerhof functional is used
as the exchange-correlation functional together with a cutoff energy
of 400 eV. The slab models are constructed with the vacuum layer thicker
than 10 \AA. For graphene/SiO$_{2}$ (MoS$_{2}$/SiO$_{2}$), the
heterostructures are constructed based on a $4\times4$ ($3\times3$)
supercell for the graphene (MoS$_{2}$) and a $2\times2$ supercell
for the SiO$_{2}$ (001) surface for better lattice match with lattice
strains smaller than 2 percent. For simulating the SiO$_{2}$ (001)
surface, a slab model with including seven Si layers is used and the
atoms in the bottom O-Si-O atomic layers are saturated with hydrogen
atoms and fixed during the structural optimization. Two types of surface
of SiO$_{2}$ with OH-rich and H-rich chemical conditions are considered
for both graphene/SiO$_{2}$ and MoS$_{2}$/SiO$_{2}$ heterostructures.
We adopt a $3\times3\times1$ Monkhorst-Pack (MP) grid for k-point
sampling for the graphene/SiO$_{2}$ and graphene/MoS$_{2}$ structures.
All the atomic models are fully relaxed until the forces are smaller
than 0.005 eV/\AA.

\begin{figure}
\includegraphics[width=8cm]{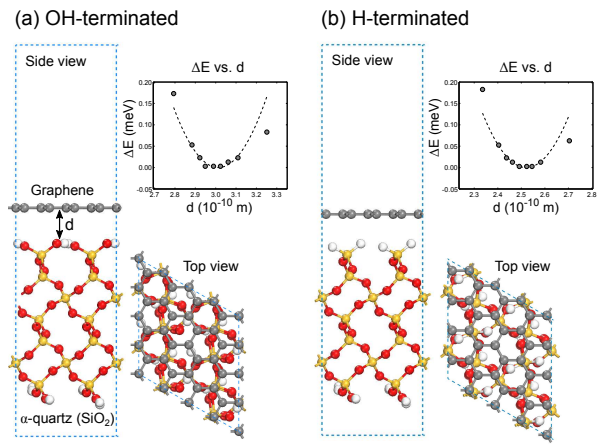}

\caption{Top and side views of the graphene/SiO$_{2}$ heterostructure used
in our density function theory (DFT) calculations to estimate $K$
at the SiO$_{2}$ (001) surface with (a) OH-termination and (b) H-termination.
The C, H, Si and O atoms are colored in gray, white, yellow and red,
respectively. The change in total energy ($\Delta E$) as a function
of $d$ and the parabolic fit used to find $K$ are also shown.}
\label{Fig:DFTStructures}
\end{figure}

Figure~\ref{Fig:DFTStructures} shows the supercell of the graphene/SiO$_{2}$
heterostructures used in our DFT calculations. The $K$ values for
the graphene/SiO$_{2}$ interface can be estimated by fitting the
parabola around the minimum of the $d-E$ curve. We also estimate
the $K$ values for the MoS$_{2}$/SiO$_{2}$ interface using the
same procedure. The extracted values ($K_{\text{OH}}$ and $K_{\text{H}}$)
are given in Table~\ref{Tab:SimParameters}. We remark that the DFT-based
estimated values of $K$ for the graphene/SiO$_{2}$ interface are
an order of magnitude than the value estimated by Cullen \emph{et
al.}~\cite{Cullen:PRL10_High} from graphene adhesion to SiO$_{2}$
($K\approx9.0\times10^{18}$ Nm$^{-3}$). Given the sensitivity of
the TBC to the numerical value of $K$, the close agreement between
our estimates of the graphene/SiO$_{2}$ TBC and experimental data
(see Table~\ref{Tab:RoomTemperatureTBCValues}) suggests that the
value of $K$ estimated in Ref.~\cite{Cullen:PRL10_High}\emph{ }is
much too low. 

\begin{table}
\begin{tabular}{|c|c|c|}
\hline 
 & Graphene & MoS$_{2}$\tabularnewline
\hline 
\hline 
$K_{\text{OH}}$ ($10^{19}$ Nm$^{-3}$) & $12.3$ & $4.94$\tabularnewline
\hline 
$K_{\text{H}}$ ($10^{19}$ Nm$^{-3}$) & $15.6$ & $2.74$\tabularnewline
\hline 
$\kappa$ (eV) & $1.1$ \cite{Persson:JPCM11_Phononic} & $9.6$ \cite{Jiang:Nanotech13_Elastic}\tabularnewline
\hline 
$\rho$ ($10^{-7}$ kgm$^{-2}$) & $7.6$  & $31$\tabularnewline
\hline 
$\alpha$  & $100$ & $100$\tabularnewline
\hline 
$\rho_{\text{sub}}$ (kgm$^{-3}$)  & \multicolumn{2}{c|}{$2200$ \cite{Persson:JPCM11_Phononic}}\tabularnewline
\hline 
$c_{L}$ (ms$^{-1}$)  & \multicolumn{2}{c|}{$5953$ \cite{Persson:JPCM11_Phononic}}\tabularnewline
\hline 
$c_{T}$ (ms$^{-1}$) & \multicolumn{2}{c|}{$3743$ \cite{Persson:JPCM11_Phononic}}\tabularnewline
\hline 
\end{tabular}

\caption{Parameters in our numerical simulations. $K_{\text{OH}}$ and $K_{\text{H}}$
are the spring constants per unit area for the OH- and H-terminated
SiO$_{2}$ interface, respectively, and also depend on the type of
2D crystal. $\kappa$ and $\rho$ are respectively the intrinsic bending
rigidity and mass density per unit area of the 2D crystal used in
Eq.~(\ref{Eq:EquationMotion_1_operator}). }
\label{Tab:SimParameters}
\end{table}

\section{Results and discussion}

We use our theory to calculate the Kapitza resistance for single-layer
graphene and MoS$_{2}$ on solid SiO$_{2}$. The parameters in our
simulations are given in Table~\ref{Tab:SimParameters}. We evaluate
the two-dimensional integral in Eq.~(\ref{Eq:TransmissionFunction})
numerically with a rectangular grid in $q$ and $\omega$. We set
a frequency cutoff of $65$ meV which approximates the maximum flexural
phonon energy at the Brillouin zone edge in most 2D materials. At
low and room temperature, the numerical value of $G$ is relatively
insensitive to the cutoff. 

We note here that the following simulated TBC values are for an idealized
flat interface at which the 2D crystal has full adhesion to the substrate
or the superstrate in the case of the encased 2D crystal. The effective
TBC values may be lower under most experimental conditions where some
surface roughness is present, resulting in the reduction of the effective
adhesion and contact area between the 2D crystal and the substrate~\cite{Persson:JCP01_Theory,Persson:EPJE10_Heat,Persson:JPCM10_Heat,Persson:JPCM11_Phononic}.
Thus, the simulated TBC values can be interpreted as the upper bounds
for the TBC of real interfaces.

\subsection{Sensitivity of numerical results to damping friction}

\begin{figure}
\includegraphics[width=7.5cm]{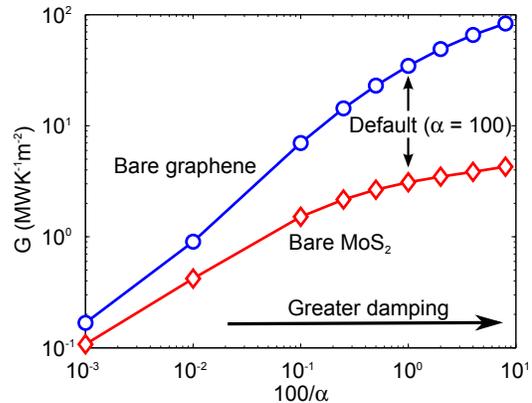}\caption{Plot of the thermal boundary conductance $G$ at 300 K at different
values of $\alpha^{-1}$ for bare graphene (circle) and MoS$_{2}$
(diamond) on OH-terminated SiO$_{2}$ surface. A decrease in the \emph{intrinsic}
damping of the flexural phonons leads to lower thermal boundary conductance
(or higher Kapitza resistance). This effect is more pronounced for
graphene.}

\label{Fig:AlphaDependenceTBC}
\end{figure}

We calculate the room temperature ($300$ K) thermal boundary conductance
for bare graphene and MoS$_{2}$ on OH-terminated SiO$_{2}$ at different
values of $\alpha$ using Eq.~(\ref{Eq:HeatTransfer_1Sheet}) and
the parameters from Table~\ref{Tab:SimParameters} except $\alpha$.
Figure~\ref{Fig:AlphaDependenceTBC} shows the TBC as a function
of $\alpha$. We find that as $\alpha$ increases and the flexural
phonon damping weakens, the TBC decreases and converges numerically
to zero. The effect is more pronounced for graphene. This indicates
that without any damping friction, the use of Eq.~(\ref{Eq:HeatTransfer_1Sheet})
yields an infinite Kapitza resistance. Physically, this means that
damping friction in single-layer 2D crystals is necessary to produce
a reasonable finite Kapitza resistance. It also confirms our earlier
analysis of the weak-coupling approximation in Ref.~\cite{Persson:JPCM11_Phononic}
which gives a TBC of $\sim300$ MWK$^{-1}$m$^{-2}$, an order of
magnitude larger than experimental data. In contrast, our default
value of $\alpha=100$ gives us a room-temperature TBC of $G=34.6$
MWK$^{-1}$m$^{-2}$ which is in very good agreement with published
experimental data of \emph{bare} single-layer graphene~\cite{Mak:APL10_Measurement,Freitag:NL09_Energy}
($\sim25$ MWK$^{-1}$m$^{-2}$) considering the uncertainty in the
input parameters. We also observe that the TBC for MoS$_{2}$ is considerably
less sensitive to the numerical value of $\alpha$ compared to graphene.
When $\alpha$ is quadrupled from $100$ to $400$, the TBC decreases
from $G=3.1$ MWK$^{-1}$m$^{-2}$ to $2.2$ MWK$^{-1}$m$^{-2}$.
This considerably weaker $\alpha$-dependence in MoS$_{2}$ implies
that our choice of $\alpha=100$ will not affect our TBC estimates
for MoS$_{2}$ significantly.

\subsection{Bare single-layer graphene and MoS$_{2}$}

\begin{figure}
\includegraphics[width=7.5cm]{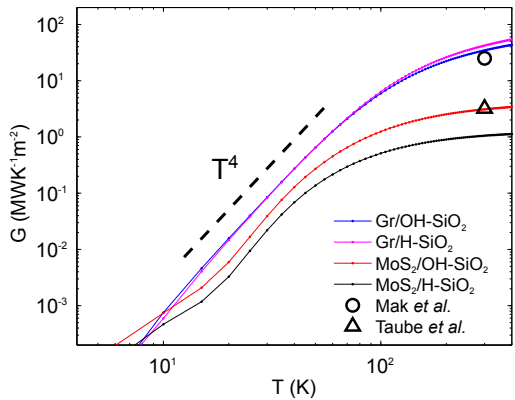}\caption{Temperature dependence of the thermal boundary conductance (TBC) for
bare single-layer graphene (Gr) and MoS$_{2}$ on OH- and H-terminated
SiO$_{2}$ surface. The data point for bare graphene from Mak \emph{et
al.}~\cite{Mak:APL10_Measurement} is represented by the open circle
while the data point for bare MoS$_{2}$ from Taube \emph{et al.}~\cite{Taube:AMI15_Temperature}
is represented by the open triangle.}
\label{Fig:TempDependTBC_Bare}
\end{figure}

We calculate the TBC for bare single-layer graphene and MoS$_{2}$
on OH- and H-terminated SiO$_{2}$ surface in the temperature range
of 5-400 K using Eq.~(\ref{Eq:HeatTransfer_1Sheet}) and the parameters
from Table~\ref{Tab:SimParameters}. The computed results are shown
in Fig.~\ref{Fig:TempDependTBC_Bare} along with the room-temperature
experimental data for bare graphene and MoS$_{2}$ from Refs.~\cite{Mak:APL10_Measurement}
and \cite{Taube:AMI15_Temperature}. At room temperature, the TBC
values for graphene are relatively close regardless of the surface
termination type (H or OH) although the H-terminated surface gives
slightly higher TBC values. In contrast, the TBC values for MoS$_{2}$
vary significantly at all temperature. At 300 K, we have $G=1.05$
and $3.10$ MWK$^{-1}$m$^{-2}$ for H- and OH-terminated SiO$_{2}$
surfaces, respectively, indicating that bare MoS$_{2}$ dissipates
heat much more efficiently to the OH-terminated SiO$_{2}$ surface
than to the H-terminated surface. This is not surprising considering
that $K_{\text{OH}}$ is significantly higher than $K_{\text{H}}$
for MoS$_{2}$ (see Table~\ref{Tab:SimParameters}) because of the
stronger coupling between MoS$_{2}$ and OH-terminated SiO$_{2}$.
The simulated room-temperature TBC values are summarized in Table~\ref{Tab:RoomTemperatureTBCValues}.
At temperatures below 100 K, the TBC scales approximately as $G\propto T^{4}$,
assuming the temperature dependence of $\gamma(\omega)\propto T$
given in Eq.~(\ref{Eq:DampingFunc}). 

We compare our simulated TBC values with those measured from experiments
on bare graphene or MoS$_{2}$ on SiO$_{2}$ at room temperature.
By using an ultrafast optical pump pulse and monitoring the transient
reflectivity on the picosecond time scale, Mak \emph{et al.}~\cite{Mak:APL10_Measurement}
obtained $\sim25$ MWK$^{-1}$m$^{-2}$ for the room-temperature TBC
of bare graphene on SiO$_{2}$. A similar value is estimated by Freitag
\emph{et al.} in Ref.~\cite{Freitag:NL09_Energy}. Ni \emph{et al.}~\cite{Ni:APL13_Few}
obtained a value of $\sim30$ MWK$^{-1}$m$^{-2}$ from their molecular
dynamics (MD) simulations of single-layer graphene on amorphous SiO$_{2}$
while Ong and Pop~\cite{Ong:PRB10_Molecular} obtained $\sim58$
MWK$^{-1}$m$^{-2}$ from the MD simulation of a single-walled carbon
nanotube on SiO$_{2}$. These experimental and MD simulation values
compare favorably with our estimate of 34.6 MWK$^{-1}$m$^{-2}$ for
graphene on OH-terminated SiO$_{2}$. For bare MoS$_{2}$ on SiO$_{2}$,
a TBC value of $G\gtrsim3.2$ MWK$^{-1}$m$^{-2}$ is estimated from
Ref.~\cite{Taube:AMI15_Temperature} using $G\gtrsim(R_{\text{tot}}-R_{\text{ox}})^{-1}$,
where $R_{\text{ox}}=1.96\times10^{-7}$ m$^{2}$KW$^{-1}$ is the
thermal resistance of 275 nm of amorphous SiO$_{2}$, with a thermal
conductivity~\cite{Regner:NComm13_Broadband} of $1.4$ MWK$^{-1}$m$^{-1}$,
and $R_{\text{tot}}=5.08\times10^{-7}$ m$^{2}$KW$^{-1}$ is the
total interfacial thermal resistance measured using an optothermal
method based on Raman spectroscopy~\cite{Taube:AMI15_Temperature}.
This estimate is remarkably close to our calculated value of $3.10$
MWK$^{-1}$m$^{-2}$ for bare MoS$_{2}$ on OH-terminated SiO$_{2}$.
For both graphene and MoS$_{2}$ , given the good agreement between
the simulated TBC values for OH-terminated SiO$_{2}$ and experimental
TBC data, this indirectly suggests that the spring constant per unit
area $K_{\text{OH}}$ for the OH-terminated surface is more representative
of real SiO$_{2}$ surface in experimental systems.

\subsection{Effect of top SiO$_{2}$ layer on thermal boundary conductance}

\begin{figure}
\includegraphics[width=7.5cm]{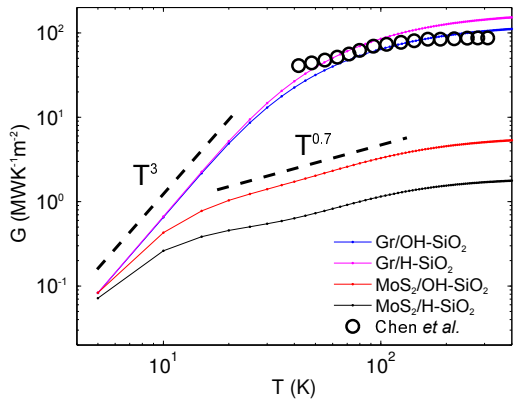}\caption{Temperature dependence of the thermal boundary conductance (TBC) for
SiO$_{2}$-encased graphene (Gr) and MoS$_{2}$ on OH- and H-terminated
SiO$_{2}$ interfaces. The open circles correspond to temperature-dependent
TBC data for single-layer encased graphene taken from Chen \emph{et
al.}~\cite{Chen:APL09_Chen}.}
\label{Fig:TempDependTBC_Encased}
\end{figure}

We calculate the TBC for SiO$_{2}$-encased single-layer graphene
and MoS$_{2}$ on OH- and H-terminated SiO$_{2}$ surface in the temperature
range of 5-400 K using Eq.~(\ref{Eq:HeatTransfer_1Sheet_TopOxide})
and the parameters from Table~\ref{Tab:SimParameters}. The computed
results are shown in Fig.~\ref{Fig:TempDependTBC_Encased} along
with the temperature-dependent experimental data for encased graphene
from Refs.~\cite{Chen:APL09_Chen} and the simulated room-temperature
TBC values are given in Table~\ref{Tab:RoomTemperatureTBCValues}.
The TBC values are significantly higher for encased graphene and MoS$_{2}$
than for the bare 2D crystals at all temperatures. In particular,
the room-temperature TBC calculated for graphene on OH-terminated
SiO$_{2}$ is $105$ MWK$^{-1}$m$^{-2}$, in good agreement with
the experimental values of $\sim83$ MWK$^{-1}$m$^{-2}$ in Ref.~\cite{Chen:APL09_Chen},
and much larger than the corresponding TBC value for bare graphene.
Like with our simulated TBC values for bare graphene, the computed
values for the OH-terminated SiO$_{2}$ interface is closer to experimental
values. However, the TBC data from Ref.~\cite{Chen:APL09_Chen} does
not decrease as much with temperature as does our simulated data.
This weaker temperature dependence may be due to the functional form
of $\gamma(\omega)$ we assumed in Eq.~(\ref{Eq:DampingFunc}). It
is possible that a more accurate calculation of the temperature-dependent
flexural phonon lifetime similar to what is done in DFT-based calculations
of the thermal conductivity~\cite{Bonini:NL12_Acoustic,Lindsay:PRB14_Phonon}
will yield a TBC temperature dependence closer to experiments. In
addition, the $\gamma(\omega)\propto T$ scaling may not hold at lower
temperatures where defect scattering of the flexural phonons may be
the limiting factor instead of anharmonic phonon interaction in real
graphene. We also find that the TBC values for MoS$_{2}$ vary significantly
at all temperature with the surface termination type. At 300 K, we
have $G=5.07$ and $1.70$ MWK$^{-1}$m$^{-2}$ for H- and OH-terminated
SiO$_{2}$ surfaces, respectively, indicating that encased MoS$_{2}$
dissipates heat much more efficiently to the OH-terminated SiO$_{2}$
surface than to the H-terminated surface like in bare MoS$_{2}$.
The temperature dependence of the TBC also varies with the 2D crystal.
At temperatures below 100 K, the TBC scales approximately as $G\propto T^{3}$
for graphene and $G\propto T^{0.7}$ for MoS$_{2}$. 

\begin{table}
\begin{tabular}{|c|c|c|c|}
\hline 
 & OH-SiO$_{2}$ & H-SiO$_{2}$ & Experimental\tabularnewline
\hline 
\hline 
Bare Graphene & $34.6$ & $42.1$ & $\sim25$ \cite{Freitag:NL09_Energy,Mak:APL10_Measurement}\tabularnewline
\hline 
Encased Graphene & $105$ & $144$ & $\sim83$ \cite{Chen:APL09_Chen}\tabularnewline
\hline 
Bare MoS$_{2}$ & $3.10$ & $1.05$ & $\gtrsim3.2$ \cite{Taube:AMI15_Temperature} \tabularnewline
\hline 
Encased MoS$_{2}$ & $5.07$ & $1.70$ & --\tabularnewline
\hline 
\end{tabular}

\caption{Computed values of the room temperature (300 K) thermal boundary conductance
(TBC) values in MWK$^{-1}$m$^{-2}$ for bare and SiO$_{2}$-encased
single-layer graphene and MoS$_{2}$ with OH- and H-terminated SiO$_{2}$
interfaces. The experimentally estimated TBC values for bare and encased
graphene and bare MoS$_{2}$ on SiO$_{2}$ are also shown for comparison.}
\label{Tab:RoomTemperatureTBCValues}
\end{table}

\subsection{Transmission analysis of bare and encased graphene}

\begin{figure}
\includegraphics[width=8cm]{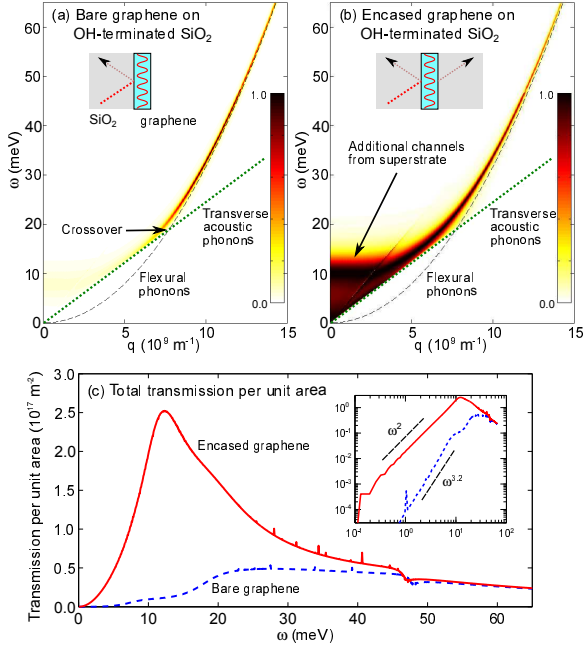}\caption{Plot of the transmission spectrum $\Xi(\mathbf{q},\omega)$ {[}see
Eq.~(\ref{Eq:TransmissionFunction}){]} for (a) bare and (b) encased
graphene on OH-terminated SiO$_{2}$ at $T=300$ K. The dashed and
dotted lines correspond to the graphene flexural ($\omega=\sqrt{\kappa/\rho}q^{2}$)
and substrate transverse acoustic ($\omega=c_{T}q$) phonon dispersion
curves, respectively. The insets show a schematic representation of
the bulk substrate phonon scattering with the graphene flexural phonon.
(c) Comparison of the total transmission per unit area spectrum $(2\pi)^{-2}\int d^{2}q\Xi(\mathbf{q},\omega)$
for bare (solid red line) and encased graphene (dashed blue line)
on OH-terminated SiO$_{2}$ at $T=300$ K. The inset shows the same
spectra but with a logarithmic scale. The small fuzzy peaks are an
artifact of the numerical integration in $q$.}
\label{Fig:TransmissionSpectrum_300K}
\end{figure}

The question the arises as to why the thermal boundary conductance
is much higher for encased 2D crystals than for their bare counterparts.
The transmission function in Eqs.~(\ref{Eq:HeatTransfer_1Sheet})
and (\ref{Eq:HeatTransfer_1Sheet_TopOxide}) is given by 
\begin{align*}
\Xi(\mathbf{q},\omega)= & \frac{4K^{2}\text{Im}D_{\text{sub}}(\mathbf{q},\omega)\text{Im}D_{\text{2D}}(\mathbf{q},\omega)}{|1-K[D_{\text{sub}}(\mathbf{q},\omega)+D_{\text{2D}}(\mathbf{q},\omega)]|^{2}}\ ,
\end{align*}
and its spectrum helps us to identify the dominant channels in heat
dissipation. For an encased 2D crystal, we use $\overline{D}_{\text{2D}}(\mathbf{q},\omega)$
instead of $D_{\text{2D}}(\mathbf{q},\omega)$ in $\Xi(\mathbf{q},\omega)$.
To illustrate the effect of the superstrate on heat dissipation, we
plot the transmission spectra for bare and encased graphene on OH-terminated
SiO$_{2}$ in Fig.~\ref{Fig:TransmissionSpectrum_300K}. Given that
$0\leq\Xi(\mathbf{q},\omega)<1$, the transmission spectrum for bare
graphene {[}see Fig.~\ref{Fig:TransmissionSpectrum_300K}(a){]} is
zero almost everywhere although the transmission peak is close to
unity along the dispersion curve for the flexural phonons for $\omega\gtrsim19$
meV, indicating the greater contribution from high-frequency modes.
At lower frequencies ($\omega<19$ meV), the transmission of individual
low-frequency modes becomes rapidly more smeared out. In addition,
the transmission is zero for $\omega<c_{T}q$ since there are no substrate
bulk acoustic phonon modes satisfying this condition, \emph{i.e.},
$\text{Im}D_{\text{sub}}(\mathbf{q},\omega)=0$ for $\omega<c_{T}q$.
Physically, this means that heat dissipation can only take place when
there is coupling to the substrate bulk acoustic phonons. Remarkably,
the contribution of the low-frequency modes to interfacial heat transfer
is relatively small for bare graphene. The crossover at $\omega=19$
meV is determined by the point where the graphene flexural phonon
dispersion curve intersects the substrate transverse acoustic phonon
dispersion curve, as can be seen in Fig.~\ref{Fig:TransmissionSpectrum_300K},
and is given by $\omega=\sqrt{\rho/\kappa}c_{T}^{2}$. Below the crossover
frequency, the substrate transverse acoustic phonons with the same
wave vector $q$ have higher energy than the corresponding graphene
flexural phonons. This reduces the probability of the flexural mode
coupling with the continuum of substrate bulk phonons, which acts
as a dissipative bath, and transferring energy into the substrate~\cite{Amorim:PRB13_Flexural}.
Another way to understand this cutoff is that the region $\omega>c_{T}q$
corresponds to substrate phonons with a transverse momentum component
of $\mathbf{q}$ and a frequency of $\omega$, and these phonons scatter
with graphene flexural phonons with the same $\mathbf{q}$ and $\omega$.
We can also associate a characteristic length scale with the crossover
point, given by $l_{c}\sim2\pi/q_{c}=0.8$ nm where $q_{c}=\sqrt{\rho/\kappa}c_{T}$.
This suggests that the TBC becomes strongly size-dependent when its
interfacial area is comparable to or smaller than $l_{c}^{2}$.

In contrast, Fig.~\ref{Fig:TransmissionSpectrum_300K}(b) shows a
very substantial low-frequency contribution in the transmissions spectrum
of encased graphene, which also can be seen in Fig.~\ref{Fig:TransmissionSpectrum_300K}(c)
where we plot the total transmission per unit area $(2\pi)^{-2}\int d^{2}q\Xi(\mathbf{q},\omega)$.
This is because the coupling of the 2D crystal to the superstrate
results in the hybridization of the 2D crystal flexural modes with
the superstrate Rayleigh modes~\cite{Amorim:PRB13_Flexural,Ong:PRB11_Effect}
and these superstrate/graphene hybrid modes can be scattered more
easily into the substrate. The enhanced transmission of these low-frequency
hybrid modes results in a significantly higher TBC for encased graphene.
This difference in the low-frequency phonon contribution in the TBC
of encased graphene also explains why the low-temperature scaling
of $G$ is different in Figs.~\ref{Fig:TempDependTBC_Bare} and \ref{Fig:TempDependTBC_Encased}.
At low frequencies, the total transmission per unit area scales as
$\omega^{3.2}$ and $\omega^{2}$ in bare and encased graphene, respectively.
The considerably weaker contribution of the low-frequency modes in
bare graphene means that at lower temperatures, the higher-frequency
modes do not contribute to interfacial heat transfer and the TBC decreases
more rapidly as the temperature is reduced. 

Physically, heat is exchanged between the SiO$_{2}$ substrate and
bare graphene when an incoming substrate bulk phonon scatters inelastically
off the graphene/SiO$_{2}$ interface and dissipates part of its energy
within the graphene through the intrinsic flexural phonon damping,
as represented schematically in the inset of Fig.~\ref{Fig:TransmissionSpectrum_300K}(a).
This scattering process efficiently dissipates energy from the substrate
to the graphene when the substrate bulk phonon has a frequency (and
hence energy) approximately equal to the corresponding graphene flexural
phonon for the same $\mathbf{q}$. On the other hand, in encased graphene,
the incoming substrate bulk phonon also scatters inelastically off
the graphene/SiO$_{2}$ interface and the energy is dissipated into
the graphene. Part of the energy is absorbed by the \emph{intrinsic}
flexural phonon damping while part of it is absorbed by the flexural
damping component from coupling to the superstrate. The latter can
be interpreted as the partial transmission of energy into the bulk
of the superstrate, as schematically represented in the inset of Fig.~\ref{Fig:TransmissionSpectrum_300K}(b). 

\subsection{Heat dissipation pathways}

\begin{figure}
\includegraphics[width=8cm]{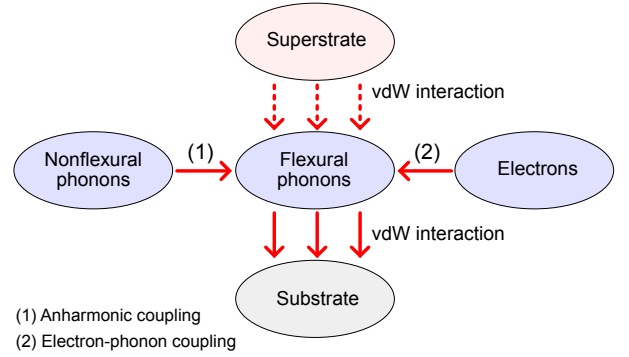}

\caption{Schematic representation of the physical model of heat dissipation
in bare and covered single-layer 2D crystals. The flexural phonons
in the bare 2D crystal absorb energy from the nonflexural phonons
and electrons via anharmonic and direct electron-phonon coupling,
and dissipate it into the substrate. The mechanical attachment of
a superstrate results in additional resonant channels for energy to
be absorbed by the flexural phonons and then transferred to the substrate,
leading to a higher thermal boundary conductance.}
\label{Fig:HeatTransferPathways}
\end{figure}

Our calculations of the thermal boundary conductance for bare and
encased single-layer 2D crystals show that the presence of a superstrate
on the 2D crystals can substantially increase the TBC between the
2D crystal and its substrate. In the bare 2D crystal, the energy from
the intrinsic degrees of freedom (nonflexural phonons and electrons)
are dissipated to the substrate via anharmonic or electron-phonon
coupling to the flexural phonons. When a superstrate is placed on
the 2D crystal, the flexural phonons are also mechanically coupled
to the Rayleigh phonon modes from the superstrate. This mechanical
coupling allows energy from the superstrate phonon modes, which we
can consider as \emph{extrinsic} degrees of freedom, to be dissipated
to the substrate. Figure~\ref{Fig:HeatTransferPathways} shows a
schematic representation of the different energy transfer pathways
involved in heat dissipation to the substrate. In our simulations
of bare and encased graphene and MoS$_{2}$, we have used a phenomenological
approach to describe the intrinsic anharmonic coupling to nonflexural
phonons. In our study, we ignore dissipation via electron-phonon coupling
because it is a second-order effect that involves two flexural phonons
and two electrons~\cite{Viljas:PRB10_Electron,Castro:PRL10_Limits,Mariani:PRL08_Flexural}
and is expected to be small although it may be significant at high
electron temperatures and densities~\cite{Song:PRL12_Disorder}.
Nevertheless, our estimates of the TBC with only dissipation from
anharmonic phonon coupling are in excellent agreement with published
experimental data, suggesting that the electron-phonon contribution
is minor. Even so, it will be an interesting exercise to determine
the TBC contribution from electron-phonon coupling at different electron
temperatures and densities.

A slightly less intuitive consequence of our proposed theoretical
framework for heat dissipation is that defect scattering of flexural
phonons in the 2D crystal may increase the TBC. Here, we invoke the
idea suggested by Song and Levitov~\cite{Song:PRL12_Disorder} that
a random distribution of short-range disorder can enlarge the scattering
phase space for electron-phonon interactions by essentially lifting
the momentum conservation restriction. Similarly, a random distribution
of defects in graphene or MoS$_{2}$ may enhance the effective anharmonic
scattering rate of flexural phonons $\gamma(\omega)$ and increase
the TBC. However, this method of enhancing the TBC may be counterproductive,
especially in graphene, from the \emph{overall} heat dissipation point
of view since the flexural phonons are the dominant thermal transport
carriers in graphene~\cite{Seol:Science10_Two,Lindsay:PRB10_Flexural}
and the decrease of the flexural phonon lifetime will lower the thermal
conductivity and reduce lateral heat dissipation.

\section{Summary and conclusions}

We have modified the theory by Persson, Volokitin and Ueba~\cite{Persson:JPCM11_Phononic}
and recast it in the Landauer form to study the Kapitza resistance
of graphene and other single-layer 2D crystals on a solid substrate.
We have shown that the weak-coupling approximation is not numerically
valid and that the inclusion of a damping function for the flexural
phonons is necessary to produce dissipation within the graphene and
yield a finite Kapitza resistance. The phenomenological form of the
damping function is deduced from published first-principles results.
We have also shown how our theory can be modified to accommodate the
effects of a superstrate. We have used DFT calculations to estimate
the spring constant per unit area for the different interfaces. Our
computed thermal boundary conductance values for bare single-layer
graphene and MoS$_{2}$ are in good agreement with published experimental
data. The theory also suggests that there is no significant contribution
to interfacial heat transfer by low-frequency phonon modes in bare
single-layer 2D crystals. 

We also find that the encasement of the 2D crystal by a SiO$_{2}$
superstrate results in a $\sim3\times$ increase in the thermal boundary
conductance for graphene. This explains why the Kapitza resistance
measured in experiments for SiO$_{2}$-encased graphene is substantially
lower than for bare graphene. This resistance reduction is also predicted
to be true for single-layer MoS$_{2}$. Our analysis shows that the
increase in the thermal boundary conductance is due to the additional
low-frequency transmission channels from coupling with the superstrate.
We also find in bare 2D crystals that there is a crossover frequency
below which phonons do not contribute significantly to cross-plane
heat dissipation to the substrate. Our calculations suggest that the
termination of the SiO$_{2}$ surface has a much stronger influence
on the Kapitza resistance for MoS$_{2}$ than for graphene. The theoretical
framework described in this work provides the basis for understanding
how the modification of interfaces and 2D crystals at the nanoscale
can be utilized to reduce the Kapitza resistance to improve interfacial
heat dissipation, an important issue in the thermal management of
nanoscale devices using 2D materials.
\begin{acknowledgments}
This work was supported in part by a grant from the Science and Engineering
Research Council (152-70-00017). We gratefully acknowledge the financial
support from the Agency for Science, Technology and Research (A{*}STAR),
Singapore and the use of computing resources at the A{*}STAR Computational
Resource Centre, Singapore. We also acknowledge discussions with Eric
Pop (Stanford University) and Justin Song (Institute of High Performance
Computing).
\end{acknowledgments}

\appendix

\section{Numerical bounds for transmission function $\Xi(\mathbf{q},\omega)$\label{Sec:NumericalBoundsTransmission}}

Let us write $KD_{\text{sub}}(\mathbf{q},\omega)=x_{1}+iy_{1}$ and
$KD_{\text{2D}}(\mathbf{q},\omega)=x_{2}+iy_{2}$ where $x_{i}$ and
$y_{i}$ are real numbers for $i=1,2$. Thus, the transmission function
in Eq.~(\ref{Eq:TransmissionFunction}) can be written as 
\[
\Xi(\mathbf{q},\omega)=\frac{4y_{1}y_{2}}{(1-x_{1}-x_{2})^{2}+(y_{1}+y_{2})^{2}}\leq\frac{4y_{1}y_{2}}{(y_{1}+y_{2})^{2}}\ .
\]
Given that $y_{1},y_{2}\geq0$, this means that 
\[
\frac{4y_{1}y_{2}}{(y_{1}+y_{2})^{2}}\leq1
\]
and therefore, we have $0\leq\Xi(\mathbf{q},\omega)\leq1$. 

\bibliographystyle{apsrev4-1}
\bibliography{references}

\begin{thebibliography}{47}%
\makeatletter
\providecommand \@ifxundefined [1]{%
 \@ifx{#1\undefined}
}%
\providecommand \@ifnum [1]{%
 \ifnum #1\expandafter \@firstoftwo
 \else \expandafter \@secondoftwo
 \fi
}%
\providecommand \@ifx [1]{%
 \ifx #1\expandafter \@firstoftwo
 \else \expandafter \@secondoftwo
 \fi
}%
\providecommand \natexlab [1]{#1}%
\providecommand \enquote  [1]{``#1''}%
\providecommand \bibnamefont  [1]{#1}%
\providecommand \bibfnamefont [1]{#1}%
\providecommand \citenamefont [1]{#1}%
\providecommand \href@noop [0]{\@secondoftwo}%
\providecommand \href [0]{\begingroup \@sanitize@url \@href}%
\providecommand \@href[1]{\@@startlink{#1}\@@href}%
\providecommand \@@href[1]{\endgroup#1\@@endlink}%
\providecommand \@sanitize@url [0]{\catcode `\\12\catcode `\$12\catcode
  `\&12\catcode `\#12\catcode `\^12\catcode `\_12\catcode `\%12\relax}%
\providecommand \@@startlink[1]{}%
\providecommand \@@endlink[0]{}%
\providecommand \url  [0]{\begingroup\@sanitize@url \@url }%
\providecommand \@url [1]{\endgroup\@href {#1}{\urlprefix }}%
\providecommand \urlprefix  [0]{URL }%
\providecommand \Eprint [0]{\href }%
\providecommand \doibase [0]{http://dx.doi.org/}%
\providecommand \selectlanguage [0]{\@gobble}%
\providecommand \bibinfo  [0]{\@secondoftwo}%
\providecommand \bibfield  [0]{\@secondoftwo}%
\providecommand \translation [1]{[#1]}%
\providecommand \BibitemOpen [0]{}%
\providecommand \bibitemStop [0]{}%
\providecommand \bibitemNoStop [0]{.\EOS\space}%
\providecommand \EOS [0]{\spacefactor3000\relax}%
\providecommand \BibitemShut  [1]{\csname bibitem#1\endcsname}%
\let\auto@bib@innerbib\@empty
\bibitem [{\citenamefont {Schwierz}(2010)}]{Schwierz:NNano10_Graphene}%
  \BibitemOpen
  \bibfield  {author} {\bibinfo {author} {\bibfnamefont {F.}~\bibnamefont
  {Schwierz}},\ }\href {\doibase 10.1038/nnano.2010.89} {\bibfield  {journal}
  {\bibinfo  {journal} {Nature Nanotechnology}\ }\textbf {\bibinfo {volume}
  {5}},\ \bibinfo {pages} {487} (\bibinfo {year} {2010})}\BibitemShut {NoStop}%
\bibitem [{\citenamefont {Wang}\ \emph {et~al.}(2012)\citenamefont {Wang},
  \citenamefont {Kalantar-Zadeh}, \citenamefont {Kis}, \citenamefont
  {Coleman},\ and\ \citenamefont {Strano}}]{Wang:NNano12_Electronics}%
  \BibitemOpen
  \bibfield  {author} {\bibinfo {author} {\bibfnamefont {Q.~H.}\ \bibnamefont
  {Wang}}, \bibinfo {author} {\bibfnamefont {K.}~\bibnamefont
  {Kalantar-Zadeh}}, \bibinfo {author} {\bibfnamefont {A.}~\bibnamefont {Kis}},
  \bibinfo {author} {\bibfnamefont {J.~N.}\ \bibnamefont {Coleman}}, \ and\
  \bibinfo {author} {\bibfnamefont {M.~S.}\ \bibnamefont {Strano}},\ }\href
  {\doibase 10.1038/nnano.2012.193} {\bibfield  {journal} {\bibinfo  {journal}
  {Nature Nanotechnology}\ }\textbf {\bibinfo {volume} {7}},\ \bibinfo {pages}
  {699} (\bibinfo {year} {2012})}\BibitemShut {NoStop}%
\bibitem [{\citenamefont {Ferrari}\ \emph {et~al.}(2015)\citenamefont
  {Ferrari}, \citenamefont {Bonaccorso}, \citenamefont {Fal'Ko}, \citenamefont
  {Novoselov}, \citenamefont {Roche}, \citenamefont {B{\o}ggild}, \citenamefont
  {Borini}, \citenamefont {Koppens}, \citenamefont {Palermo}, \citenamefont
  {Pugno} \emph {et~al.}}]{Ferrari:Nanoscale15_Science}%
  \BibitemOpen
  \bibfield  {author} {\bibinfo {author} {\bibfnamefont {A.~C.}\ \bibnamefont
  {Ferrari}}, \bibinfo {author} {\bibfnamefont {F.}~\bibnamefont {Bonaccorso}},
  \bibinfo {author} {\bibfnamefont {V.}~\bibnamefont {Fal'Ko}}, \bibinfo
  {author} {\bibfnamefont {K.~S.}\ \bibnamefont {Novoselov}}, \bibinfo {author}
  {\bibfnamefont {S.}~\bibnamefont {Roche}}, \bibinfo {author} {\bibfnamefont
  {P.}~\bibnamefont {B{\o}ggild}}, \bibinfo {author} {\bibfnamefont
  {S.}~\bibnamefont {Borini}}, \bibinfo {author} {\bibfnamefont {F.~H.}\
  \bibnamefont {Koppens}}, \bibinfo {author} {\bibfnamefont {V.}~\bibnamefont
  {Palermo}}, \bibinfo {author} {\bibfnamefont {N.}~\bibnamefont {Pugno}},
  \emph {et~al.},\ }\href {\doibase 10.1039/c4nr01600a} {\bibfield  {journal}
  {\bibinfo  {journal} {Nanoscale}\ }\textbf {\bibinfo {volume} {7}},\ \bibinfo
  {pages} {4598} (\bibinfo {year} {2015})}\BibitemShut {NoStop}%
\bibitem [{\citenamefont {Liu}\ \emph {et~al.}(2012)\citenamefont {Liu},
  \citenamefont {Neal},\ and\ \citenamefont {Ye}}]{Liu:ACSNano12_Channel}%
  \BibitemOpen
  \bibfield  {author} {\bibinfo {author} {\bibfnamefont {H.}~\bibnamefont
  {Liu}}, \bibinfo {author} {\bibfnamefont {A.~T.}\ \bibnamefont {Neal}}, \
  and\ \bibinfo {author} {\bibfnamefont {P.~D.}\ \bibnamefont {Ye}},\ }\href
  {\doibase 10.1021/nn303513c} {\bibfield  {journal} {\bibinfo  {journal} {ACS
  Nano}\ }\textbf {\bibinfo {volume} {6}},\ \bibinfo {pages} {8563} (\bibinfo
  {year} {2012})}\BibitemShut {NoStop}%
\bibitem [{\citenamefont {Lembke}\ \emph {et~al.}(2015)\citenamefont {Lembke},
  \citenamefont {Bertolazzi},\ and\ \citenamefont {Kis}}]{Lembke:ACR15_Single}%
  \BibitemOpen
  \bibfield  {author} {\bibinfo {author} {\bibfnamefont {D.}~\bibnamefont
  {Lembke}}, \bibinfo {author} {\bibfnamefont {S.}~\bibnamefont {Bertolazzi}},
  \ and\ \bibinfo {author} {\bibfnamefont {A.}~\bibnamefont {Kis}},\ }\href
  {\doibase 10.1021/ar500274q} {\bibfield  {journal} {\bibinfo  {journal}
  {Accounts of Chemical Research}\ }\textbf {\bibinfo {volume} {48}},\ \bibinfo
  {pages} {100} (\bibinfo {year} {2015})}\BibitemShut {NoStop}%
\bibitem [{\citenamefont {Meric}\ \emph {et~al.}(2008)\citenamefont {Meric},
  \citenamefont {Han}, \citenamefont {Young}, \citenamefont {Ozyilmaz},
  \citenamefont {Kim},\ and\ \citenamefont {Shepard}}]{Meric:NNano08_Current}%
  \BibitemOpen
  \bibfield  {author} {\bibinfo {author} {\bibfnamefont {I.}~\bibnamefont
  {Meric}}, \bibinfo {author} {\bibfnamefont {M.~Y.}\ \bibnamefont {Han}},
  \bibinfo {author} {\bibfnamefont {A.~F.}\ \bibnamefont {Young}}, \bibinfo
  {author} {\bibfnamefont {B.}~\bibnamefont {Ozyilmaz}}, \bibinfo {author}
  {\bibfnamefont {P.}~\bibnamefont {Kim}}, \ and\ \bibinfo {author}
  {\bibfnamefont {K.~L.}\ \bibnamefont {Shepard}},\ }\href {\doibase
  10.1038/nnano.2008.268} {\bibfield  {journal} {\bibinfo  {journal} {Nature
  Nanotechnology}\ }\textbf {\bibinfo {volume} {3}},\ \bibinfo {pages} {654}
  (\bibinfo {year} {2008})}\BibitemShut {NoStop}%
\bibitem [{\citenamefont {Serov}\ \emph {et~al.}(2014)\citenamefont {Serov},
  \citenamefont {Ong}, \citenamefont {Fischetti},\ and\ \citenamefont
  {Pop}}]{Serov:JAP14_Theoretical}%
  \BibitemOpen
  \bibfield  {author} {\bibinfo {author} {\bibfnamefont {A.~Y.}\ \bibnamefont
  {Serov}}, \bibinfo {author} {\bibfnamefont {Z.-Y.}\ \bibnamefont {Ong}},
  \bibinfo {author} {\bibfnamefont {M.~V.}\ \bibnamefont {Fischetti}}, \ and\
  \bibinfo {author} {\bibfnamefont {E.}~\bibnamefont {Pop}},\ }\href {\doibase
  10.1063/1.4884614} {\bibfield  {journal} {\bibinfo  {journal} {J. Appl.
  Phys.}\ }\textbf {\bibinfo {volume} {116}},\ \bibinfo {pages} {034507}
  (\bibinfo {year} {2014})}\BibitemShut {NoStop}%
\bibitem [{\citenamefont {Pop}(2010)}]{Pop:NResearch10_Energy}%
  \BibitemOpen
  \bibfield  {author} {\bibinfo {author} {\bibfnamefont {E.}~\bibnamefont
  {Pop}},\ }\href {\doibase 10.1007/s12274-010-1019-z} {\bibfield  {journal}
  {\bibinfo  {journal} {Nano Research}\ }\textbf {\bibinfo {volume} {3}},\
  \bibinfo {pages} {147} (\bibinfo {year} {2010})}\BibitemShut {NoStop}%
\bibitem [{\citenamefont {Bae}\ \emph {et~al.}(2010)\citenamefont {Bae},
  \citenamefont {Ong}, \citenamefont {Estrada},\ and\ \citenamefont
  {Pop}}]{Bae:NL10_Imaging}%
  \BibitemOpen
  \bibfield  {author} {\bibinfo {author} {\bibfnamefont {M.-H.}\ \bibnamefont
  {Bae}}, \bibinfo {author} {\bibfnamefont {Z.-Y.}\ \bibnamefont {Ong}},
  \bibinfo {author} {\bibfnamefont {D.}~\bibnamefont {Estrada}}, \ and\
  \bibinfo {author} {\bibfnamefont {E.}~\bibnamefont {Pop}},\ }\href {\doibase
  10.1021/nl1011596} {\bibfield  {journal} {\bibinfo  {journal} {Nano Lett.}\
  }\textbf {\bibinfo {volume} {10}},\ \bibinfo {pages} {4787} (\bibinfo {year}
  {2010})}\BibitemShut {NoStop}%
\bibitem [{\citenamefont {Volokitin}\ and\ \citenamefont
  {Persson}(2011)}]{Volokitin:PRB11_Near}%
  \BibitemOpen
  \bibfield  {author} {\bibinfo {author} {\bibfnamefont {A.}~\bibnamefont
  {Volokitin}}\ and\ \bibinfo {author} {\bibfnamefont {B.}~\bibnamefont
  {Persson}},\ }\href {\doibase 10.1103/PhysRevB.83.241407} {\bibfield
  {journal} {\bibinfo  {journal} {Phys. Rev. B}\ }\textbf {\bibinfo {volume}
  {83}},\ \bibinfo {pages} {241407} (\bibinfo {year} {2011})}\BibitemShut
  {NoStop}%
\bibitem [{\citenamefont {Peng}\ \emph {et~al.}(2015)\citenamefont {Peng},
  \citenamefont {Zhang},\ and\ \citenamefont {Li}}]{Peng:APL15_Thermal}%
  \BibitemOpen
  \bibfield  {author} {\bibinfo {author} {\bibfnamefont {J.}~\bibnamefont
  {Peng}}, \bibinfo {author} {\bibfnamefont {G.}~\bibnamefont {Zhang}}, \ and\
  \bibinfo {author} {\bibfnamefont {B.}~\bibnamefont {Li}},\ }\href {\doibase
  10.1063/1.4932125} {\bibfield  {journal} {\bibinfo  {journal} {Appl. Phys.
  Lett.}\ }\textbf {\bibinfo {volume} {107}},\ \bibinfo {pages} {133108}
  (\bibinfo {year} {2015})}\BibitemShut {NoStop}%
\bibitem [{\citenamefont {Ong}\ \emph {et~al.}(2013)\citenamefont {Ong},
  \citenamefont {Fischetti}, \citenamefont {Serov},\ and\ \citenamefont
  {Pop}}]{Ong:PRB13_Signature}%
  \BibitemOpen
  \bibfield  {author} {\bibinfo {author} {\bibfnamefont {Z.-Y.}\ \bibnamefont
  {Ong}}, \bibinfo {author} {\bibfnamefont {M.~V.}\ \bibnamefont {Fischetti}},
  \bibinfo {author} {\bibfnamefont {A.~Y.}\ \bibnamefont {Serov}}, \ and\
  \bibinfo {author} {\bibfnamefont {E.}~\bibnamefont {Pop}},\ }\href {\doibase
  10.1103/PhysRevB.87.195404} {\bibfield  {journal} {\bibinfo  {journal} {Phys.
  Rev. B}\ }\textbf {\bibinfo {volume} {87}},\ \bibinfo {pages} {195404}
  (\bibinfo {year} {2013})}\BibitemShut {NoStop}%
\bibitem [{\citenamefont {Persson}\ \emph {et~al.}(2011)\citenamefont
  {Persson}, \citenamefont {Volokitin},\ and\ \citenamefont
  {Ueba}}]{Persson:JPCM11_Phononic}%
  \BibitemOpen
  \bibfield  {author} {\bibinfo {author} {\bibfnamefont {B.~N.~J.}\
  \bibnamefont {Persson}}, \bibinfo {author} {\bibfnamefont {A.~I.}\
  \bibnamefont {Volokitin}}, \ and\ \bibinfo {author} {\bibfnamefont
  {H.}~\bibnamefont {Ueba}},\ }\href {\doibase 10.1088/0953-8984/23/4/045009}
  {\bibfield  {journal} {\bibinfo  {journal} {J. Phys.: Condens. Matter}\
  }\textbf {\bibinfo {volume} {23}},\ \bibinfo {pages} {045009} (\bibinfo
  {year} {2011})}\BibitemShut {NoStop}%
\bibitem [{\citenamefont {Chen}\ \emph {et~al.}(2009)\citenamefont {Chen},
  \citenamefont {Jang}, \citenamefont {Bao}, \citenamefont {Lau},\ and\
  \citenamefont {Dames}}]{Chen:APL09_Chen}%
  \BibitemOpen
  \bibfield  {author} {\bibinfo {author} {\bibfnamefont {Z.}~\bibnamefont
  {Chen}}, \bibinfo {author} {\bibfnamefont {W.}~\bibnamefont {Jang}}, \bibinfo
  {author} {\bibfnamefont {W.}~\bibnamefont {Bao}}, \bibinfo {author}
  {\bibfnamefont {C.~N.}\ \bibnamefont {Lau}}, \ and\ \bibinfo {author}
  {\bibfnamefont {C.}~\bibnamefont {Dames}},\ }\href {\doibase
  10.1063/1.3245315} {\bibfield  {journal} {\bibinfo  {journal} {Appl. Phys.
  Lett.}\ }\textbf {\bibinfo {volume} {95}},\ \bibinfo {pages} {161910}
  (\bibinfo {year} {2009})}\BibitemShut {NoStop}%
\bibitem [{\citenamefont {Mak}\ \emph {et~al.}(2010)\citenamefont {Mak},
  \citenamefont {Lui},\ and\ \citenamefont {Heinz}}]{Mak:APL10_Measurement}%
  \BibitemOpen
  \bibfield  {author} {\bibinfo {author} {\bibfnamefont {K.~F.}\ \bibnamefont
  {Mak}}, \bibinfo {author} {\bibfnamefont {C.~H.}\ \bibnamefont {Lui}}, \ and\
  \bibinfo {author} {\bibfnamefont {T.~F.}\ \bibnamefont {Heinz}},\ }\href
  {\doibase 10.1063/1.3511537} {\bibfield  {journal} {\bibinfo  {journal}
  {Appl. Phys. Lett.}\ }\textbf {\bibinfo {volume} {97}},\ \bibinfo {pages}
  {221904} (\bibinfo {year} {2010})}\BibitemShut {NoStop}%
\bibitem [{\citenamefont {Ong}\ and\ \citenamefont
  {Pop}(2011)}]{Ong:PRB11_Effect}%
  \BibitemOpen
  \bibfield  {author} {\bibinfo {author} {\bibfnamefont {Z.-Y.}\ \bibnamefont
  {Ong}}\ and\ \bibinfo {author} {\bibfnamefont {E.}~\bibnamefont {Pop}},\
  }\href {\doibase 10.1103/PhysRevB.84.075471} {\bibfield  {journal} {\bibinfo
  {journal} {Phys. Rev. B}\ }\textbf {\bibinfo {volume} {84}},\ \bibinfo
  {pages} {075471} (\bibinfo {year} {2011})}\BibitemShut {NoStop}%
\bibitem [{\citenamefont {Swartz}\ and\ \citenamefont
  {Pohl}(1989)}]{Swartz:RMP89_Thermal}%
  \BibitemOpen
  \bibfield  {author} {\bibinfo {author} {\bibfnamefont {E.~T.}\ \bibnamefont
  {Swartz}}\ and\ \bibinfo {author} {\bibfnamefont {R.~O.}\ \bibnamefont
  {Pohl}},\ }\href {\doibase 10.1103/RevModPhys.61.605} {\bibfield  {journal}
  {\bibinfo  {journal} {Rev. Mod. Phys.}\ }\textbf {\bibinfo {volume} {61}},\
  \bibinfo {pages} {605} (\bibinfo {year} {1989})}\BibitemShut {NoStop}%
\bibitem [{\citenamefont {Freitag}\ \emph {et~al.}(2009)\citenamefont
  {Freitag}, \citenamefont {Steiner}, \citenamefont {Martin}, \citenamefont
  {Perebeinos}, \citenamefont {Chen}, \citenamefont {Tsang},\ and\
  \citenamefont {Avouris}}]{Freitag:NL09_Energy}%
  \BibitemOpen
  \bibfield  {author} {\bibinfo {author} {\bibfnamefont {M.}~\bibnamefont
  {Freitag}}, \bibinfo {author} {\bibfnamefont {M.}~\bibnamefont {Steiner}},
  \bibinfo {author} {\bibfnamefont {Y.}~\bibnamefont {Martin}}, \bibinfo
  {author} {\bibfnamefont {V.}~\bibnamefont {Perebeinos}}, \bibinfo {author}
  {\bibfnamefont {Z.}~\bibnamefont {Chen}}, \bibinfo {author} {\bibfnamefont
  {J.~C.}\ \bibnamefont {Tsang}}, \ and\ \bibinfo {author} {\bibfnamefont
  {P.}~\bibnamefont {Avouris}},\ }\href {\doibase 10.1021/nl803883h} {\bibfield
   {journal} {\bibinfo  {journal} {Nano Lett.}\ }\textbf {\bibinfo {volume}
  {9}},\ \bibinfo {pages} {1883} (\bibinfo {year} {2009})}\BibitemShut
  {NoStop}%
\bibitem [{\citenamefont {Taube}\ \emph {et~al.}(2015)\citenamefont {Taube},
  \citenamefont {Judek}, \citenamefont {Lapinska},\ and\ \citenamefont
  {Zdrojek}}]{Taube:AMI15_Temperature}%
  \BibitemOpen
  \bibfield  {author} {\bibinfo {author} {\bibfnamefont {A.}~\bibnamefont
  {Taube}}, \bibinfo {author} {\bibfnamefont {J.}~\bibnamefont {Judek}},
  \bibinfo {author} {\bibfnamefont {A.}~\bibnamefont {Lapinska}}, \ and\
  \bibinfo {author} {\bibfnamefont {M.}~\bibnamefont {Zdrojek}},\ }\href
  {\doibase 10.1021/acsami.5b00690} {\bibfield  {journal} {\bibinfo  {journal}
  {Appl. Mater. Interfaces}\ }\textbf {\bibinfo {volume} {7}},\ \bibinfo
  {pages} {5061} (\bibinfo {year} {2015})}\BibitemShut {NoStop}%
\bibitem [{\citenamefont {Ni}\ \emph {et~al.}(2013)\citenamefont {Ni},
  \citenamefont {Chalopin},\ and\ \citenamefont {Volz}}]{Ni:APL13_Few}%
  \BibitemOpen
  \bibfield  {author} {\bibinfo {author} {\bibfnamefont {Y.}~\bibnamefont
  {Ni}}, \bibinfo {author} {\bibfnamefont {Y.}~\bibnamefont {Chalopin}}, \ and\
  \bibinfo {author} {\bibfnamefont {S.}~\bibnamefont {Volz}},\ }\href {\doibase
  10.1063/1.4824013} {\bibfield  {journal} {\bibinfo  {journal} {Appl. Phys.
  Lett.}\ }\textbf {\bibinfo {volume} {103}},\ \bibinfo {pages} {141905}
  (\bibinfo {year} {2013})}\BibitemShut {NoStop}%
\bibitem [{\citenamefont {Amorim}\ and\ \citenamefont
  {Guinea}(2013)}]{Amorim:PRB13_Flexural}%
  \BibitemOpen
  \bibfield  {author} {\bibinfo {author} {\bibfnamefont {B.}~\bibnamefont
  {Amorim}}\ and\ \bibinfo {author} {\bibfnamefont {F.}~\bibnamefont
  {Guinea}},\ }\href {\doibase 10.1103/PhysRevB.88.115418} {\bibfield
  {journal} {\bibinfo  {journal} {Phys. Rev. B}\ }\textbf {\bibinfo {volume}
  {88}},\ \bibinfo {pages} {115418} (\bibinfo {year} {2013})}\BibitemShut
  {NoStop}%
\bibitem [{\citenamefont {Persson}(2001)}]{Persson:JCP01_Theory}%
  \BibitemOpen
  \bibfield  {author} {\bibinfo {author} {\bibfnamefont {B.~N.~J.}\
  \bibnamefont {Persson}},\ }\href {\doibase 10.1063/1.1388626} {\bibfield
  {journal} {\bibinfo  {journal} {J. Chem. Phys.}\ }\textbf {\bibinfo {volume}
  {115}},\ \bibinfo {pages} {3840} (\bibinfo {year} {2001})}\BibitemShut
  {NoStop}%
\bibitem [{\citenamefont {Fallahazad}\ \emph {et~al.}(2010)\citenamefont
  {Fallahazad}, \citenamefont {Kim}, \citenamefont {Colombo},\ and\
  \citenamefont {Tutuc}}]{Fallahazad:APL10_Dielectric}%
  \BibitemOpen
  \bibfield  {author} {\bibinfo {author} {\bibfnamefont {B.}~\bibnamefont
  {Fallahazad}}, \bibinfo {author} {\bibfnamefont {S.}~\bibnamefont {Kim}},
  \bibinfo {author} {\bibfnamefont {L.}~\bibnamefont {Colombo}}, \ and\
  \bibinfo {author} {\bibfnamefont {E.}~\bibnamefont {Tutuc}},\ }\href
  {\doibase 10.1063/1.3492843} {\bibfield  {journal} {\bibinfo  {journal}
  {Appl. Phys. Lett.}\ }\textbf {\bibinfo {volume} {97}},\ \bibinfo {pages}
  {123105} (\bibinfo {year} {2010})}\BibitemShut {NoStop}%
\bibitem [{\citenamefont {Fallahazad}\ \emph {et~al.}(2012)\citenamefont
  {Fallahazad}, \citenamefont {Lee}, \citenamefont {Lian}, \citenamefont {Kim},
  \citenamefont {Corbet}, \citenamefont {Ferrer}, \citenamefont {Colombo},\
  and\ \citenamefont {Tutuc}}]{Fallahazad:APL12_Scaling}%
  \BibitemOpen
  \bibfield  {author} {\bibinfo {author} {\bibfnamefont {B.}~\bibnamefont
  {Fallahazad}}, \bibinfo {author} {\bibfnamefont {K.}~\bibnamefont {Lee}},
  \bibinfo {author} {\bibfnamefont {G.}~\bibnamefont {Lian}}, \bibinfo {author}
  {\bibfnamefont {S.}~\bibnamefont {Kim}}, \bibinfo {author} {\bibfnamefont
  {C.}~\bibnamefont {Corbet}}, \bibinfo {author} {\bibfnamefont
  {D.}~\bibnamefont {Ferrer}}, \bibinfo {author} {\bibfnamefont
  {L.}~\bibnamefont {Colombo}}, \ and\ \bibinfo {author} {\bibfnamefont
  {E.}~\bibnamefont {Tutuc}},\ }\href {\doibase 10.1063/1.3689785} {\bibfield
  {journal} {\bibinfo  {journal} {Appl. Phys. Lett.}\ }\textbf {\bibinfo
  {volume} {100}},\ \bibinfo {pages} {093112} (\bibinfo {year}
  {2012})}\BibitemShut {NoStop}%
\bibitem [{\citenamefont {Ong}\ and\ \citenamefont
  {Fischetti}(2012)}]{Ong:PRB12_Charged}%
  \BibitemOpen
  \bibfield  {author} {\bibinfo {author} {\bibfnamefont {Z.-Y.}\ \bibnamefont
  {Ong}}\ and\ \bibinfo {author} {\bibfnamefont {M.~V.}\ \bibnamefont
  {Fischetti}},\ }\href {\doibase 10.1103/PhysRevB.86.121409} {\bibfield
  {journal} {\bibinfo  {journal} {Phys. Rev. B}\ }\textbf {\bibinfo {volume}
  {86}},\ \bibinfo {pages} {121409} (\bibinfo {year} {2012})}\BibitemShut
  {NoStop}%
\bibitem [{\citenamefont {Ong}\ and\ \citenamefont
  {Fischetti}(2013{\natexlab{a}})}]{Ong:APL13_Top}%
  \BibitemOpen
  \bibfield  {author} {\bibinfo {author} {\bibfnamefont {Z.-Y.}\ \bibnamefont
  {Ong}}\ and\ \bibinfo {author} {\bibfnamefont {M.~V.}\ \bibnamefont
  {Fischetti}},\ }\href {\doibase 10.1063/1.4804432} {\bibfield  {journal}
  {\bibinfo  {journal} {Appl. Phys. Lett.}\ }\textbf {\bibinfo {volume}
  {102}},\ \bibinfo {pages} {183506} (\bibinfo {year}
  {2013}{\natexlab{a}})}\BibitemShut {NoStop}%
\bibitem [{\citenamefont {Radisavljevic}\ and\ \citenamefont
  {Kis}(2013)}]{Radisavljevic:NMat13_Mobility}%
  \BibitemOpen
  \bibfield  {author} {\bibinfo {author} {\bibfnamefont {B.}~\bibnamefont
  {Radisavljevic}}\ and\ \bibinfo {author} {\bibfnamefont {A.}~\bibnamefont
  {Kis}},\ }\href {\doibase 10.1038/nmat3687} {\bibfield  {journal} {\bibinfo
  {journal} {Nature Materials}\ }\textbf {\bibinfo {volume} {12}},\ \bibinfo
  {pages} {815} (\bibinfo {year} {2013})}\BibitemShut {NoStop}%
\bibitem [{\citenamefont {Ong}\ and\ \citenamefont
  {Fischetti}(2013{\natexlab{b}})}]{Ong:PRB13_Mobility}%
  \BibitemOpen
  \bibfield  {author} {\bibinfo {author} {\bibfnamefont {Z.-Y.}\ \bibnamefont
  {Ong}}\ and\ \bibinfo {author} {\bibfnamefont {M.~V.}\ \bibnamefont
  {Fischetti}},\ }\href {\doibase 10.1103/PhysRevB.88.165316} {\bibfield
  {journal} {\bibinfo  {journal} {Phys. Rev. B}\ }\textbf {\bibinfo {volume}
  {88}},\ \bibinfo {pages} {165316} (\bibinfo {year}
  {2013}{\natexlab{b}})}\BibitemShut {NoStop}%
\bibitem [{\citenamefont {Bonini}\ \emph {et~al.}(2012)\citenamefont {Bonini},
  \citenamefont {Garg},\ and\ \citenamefont {Marzari}}]{Bonini:NL12_Acoustic}%
  \BibitemOpen
  \bibfield  {author} {\bibinfo {author} {\bibfnamefont {N.}~\bibnamefont
  {Bonini}}, \bibinfo {author} {\bibfnamefont {J.}~\bibnamefont {Garg}}, \ and\
  \bibinfo {author} {\bibfnamefont {N.}~\bibnamefont {Marzari}},\ }\href
  {\doibase 10.1021/nl202694m} {\bibfield  {journal} {\bibinfo  {journal} {Nano
  Lett.}\ }\textbf {\bibinfo {volume} {12}},\ \bibinfo {pages} {2673} (\bibinfo
  {year} {2012})}\BibitemShut {NoStop}%
\bibitem [{\citenamefont {Lindsay}\ \emph {et~al.}(2014)\citenamefont
  {Lindsay}, \citenamefont {Li}, \citenamefont {Carrete}, \citenamefont
  {Mingo}, \citenamefont {Broido},\ and\ \citenamefont
  {Reinecke}}]{Lindsay:PRB14_Phonon}%
  \BibitemOpen
  \bibfield  {author} {\bibinfo {author} {\bibfnamefont {L.}~\bibnamefont
  {Lindsay}}, \bibinfo {author} {\bibfnamefont {W.}~\bibnamefont {Li}},
  \bibinfo {author} {\bibfnamefont {J.}~\bibnamefont {Carrete}}, \bibinfo
  {author} {\bibfnamefont {N.}~\bibnamefont {Mingo}}, \bibinfo {author}
  {\bibfnamefont {D.~A.}\ \bibnamefont {Broido}}, \ and\ \bibinfo {author}
  {\bibfnamefont {T.~L.}\ \bibnamefont {Reinecke}},\ }\href {\doibase
  10.1103/PhysRevB.89.155426} {\bibfield  {journal} {\bibinfo  {journal} {Phys.
  Rev. B}\ }\textbf {\bibinfo {volume} {89}},\ \bibinfo {pages} {155426}
  (\bibinfo {year} {2014})}\BibitemShut {NoStop}%
\bibitem [{\citenamefont {Viljas}\ and\ \citenamefont
  {Heikkil\"a}(2010)}]{Viljas:PRB10_Electron}%
  \BibitemOpen
  \bibfield  {author} {\bibinfo {author} {\bibfnamefont {J.~K.}\ \bibnamefont
  {Viljas}}\ and\ \bibinfo {author} {\bibfnamefont {T.~T.}\ \bibnamefont
  {Heikkil\"a}},\ }\href {\doibase 10.1103/PhysRevB.81.245404} {\bibfield
  {journal} {\bibinfo  {journal} {Phys. Rev. B}\ }\textbf {\bibinfo {volume}
  {81}},\ \bibinfo {pages} {245404} (\bibinfo {year} {2010})}\BibitemShut
  {NoStop}%
\bibitem [{\citenamefont {Castro}\ \emph {et~al.}(2010)\citenamefont {Castro},
  \citenamefont {Ochoa}, \citenamefont {Katsnelson}, \citenamefont {Gorbachev},
  \citenamefont {Elias}, \citenamefont {Novoselov}, \citenamefont {Geim},\ and\
  \citenamefont {Guinea}}]{Castro:PRL10_Limits}%
  \BibitemOpen
  \bibfield  {author} {\bibinfo {author} {\bibfnamefont {E.~V.}\ \bibnamefont
  {Castro}}, \bibinfo {author} {\bibfnamefont {H.}~\bibnamefont {Ochoa}},
  \bibinfo {author} {\bibfnamefont {M.~I.}\ \bibnamefont {Katsnelson}},
  \bibinfo {author} {\bibfnamefont {R.~V.}\ \bibnamefont {Gorbachev}}, \bibinfo
  {author} {\bibfnamefont {D.~C.}\ \bibnamefont {Elias}}, \bibinfo {author}
  {\bibfnamefont {K.~S.}\ \bibnamefont {Novoselov}}, \bibinfo {author}
  {\bibfnamefont {A.~K.}\ \bibnamefont {Geim}}, \ and\ \bibinfo {author}
  {\bibfnamefont {F.}~\bibnamefont {Guinea}},\ }\href {\doibase
  10.1103/PhysRevLett.105.266601} {\bibfield  {journal} {\bibinfo  {journal}
  {Phys. Rev. Lett.}\ }\textbf {\bibinfo {volume} {105}},\ \bibinfo {pages}
  {266601} (\bibinfo {year} {2010})}\BibitemShut {NoStop}%
\bibitem [{\citenamefont {Mariani}\ and\ \citenamefont {von
  Oppen}(2008)}]{Mariani:PRL08_Flexural}%
  \BibitemOpen
  \bibfield  {author} {\bibinfo {author} {\bibfnamefont {E.}~\bibnamefont
  {Mariani}}\ and\ \bibinfo {author} {\bibfnamefont {F.}~\bibnamefont {von
  Oppen}},\ }\href {\doibase 10.1103/PhysRevLett.100.076801} {\bibfield
  {journal} {\bibinfo  {journal} {Phys. Rev. Lett.}\ }\textbf {\bibinfo
  {volume} {100}},\ \bibinfo {pages} {076801} (\bibinfo {year}
  {2008})}\BibitemShut {NoStop}%
\bibitem [{\citenamefont {Lindsay}\ \emph {et~al.}(2010)\citenamefont
  {Lindsay}, \citenamefont {Broido},\ and\ \citenamefont
  {Mingo}}]{Lindsay:PRB10_Flexural}%
  \BibitemOpen
  \bibfield  {author} {\bibinfo {author} {\bibfnamefont {L.}~\bibnamefont
  {Lindsay}}, \bibinfo {author} {\bibfnamefont {D.~A.}\ \bibnamefont {Broido}},
  \ and\ \bibinfo {author} {\bibfnamefont {N.}~\bibnamefont {Mingo}},\ }\href
  {\doibase 10.1103/PhysRevB.82.115427} {\bibfield  {journal} {\bibinfo
  {journal} {Phys. Rev. B}\ }\textbf {\bibinfo {volume} {82}},\ \bibinfo
  {pages} {115427} (\bibinfo {year} {2010})}\BibitemShut {NoStop}%
\bibitem [{\citenamefont {Li}\ \emph {et~al.}(2013)\citenamefont {Li},
  \citenamefont {Carrete},\ and\ \citenamefont {Mingo}}]{Li:APL13_Thermal}%
  \BibitemOpen
  \bibfield  {author} {\bibinfo {author} {\bibfnamefont {W.}~\bibnamefont
  {Li}}, \bibinfo {author} {\bibfnamefont {J.}~\bibnamefont {Carrete}}, \ and\
  \bibinfo {author} {\bibfnamefont {N.}~\bibnamefont {Mingo}},\ }\href
  {\doibase 10.1063/1.4850995} {\bibfield  {journal} {\bibinfo  {journal}
  {Appl. Phys. Lett.}\ }\textbf {\bibinfo {volume} {103}},\ \bibinfo {pages}
  {253103} (\bibinfo {year} {2013})}\BibitemShut {NoStop}%
\bibitem [{\citenamefont {Wei}\ \emph {et~al.}(2014)\citenamefont {Wei},
  \citenamefont {Wang}, \citenamefont {Shen}, \citenamefont {Xie},
  \citenamefont {Xiao}, \citenamefont {Zhong},\ and\ \citenamefont
  {Zhang}}]{Wei:APL14_Phonon}%
  \BibitemOpen
  \bibfield  {author} {\bibinfo {author} {\bibfnamefont {X.}~\bibnamefont
  {Wei}}, \bibinfo {author} {\bibfnamefont {Y.}~\bibnamefont {Wang}}, \bibinfo
  {author} {\bibfnamefont {Y.}~\bibnamefont {Shen}}, \bibinfo {author}
  {\bibfnamefont {G.}~\bibnamefont {Xie}}, \bibinfo {author} {\bibfnamefont
  {H.}~\bibnamefont {Xiao}}, \bibinfo {author} {\bibfnamefont {J.}~\bibnamefont
  {Zhong}}, \ and\ \bibinfo {author} {\bibfnamefont {G.}~\bibnamefont
  {Zhang}},\ }\href {\doibase 10.1063/1.4895344} {\bibfield  {journal}
  {\bibinfo  {journal} {Appl. Phys. Lett.}\ }\textbf {\bibinfo {volume}
  {105}},\ \bibinfo {pages} {103902} (\bibinfo {year} {2014})}\BibitemShut
  {NoStop}%
\bibitem [{\citenamefont {Jiang}\ \emph {et~al.}(2014)\citenamefont {Jiang},
  \citenamefont {Park},\ and\ \citenamefont
  {Rabczuk}}]{Jiang:Nanoscale14_MoS2}%
  \BibitemOpen
  \bibfield  {author} {\bibinfo {author} {\bibfnamefont {J.-W.}\ \bibnamefont
  {Jiang}}, \bibinfo {author} {\bibfnamefont {H.~S.}\ \bibnamefont {Park}}, \
  and\ \bibinfo {author} {\bibfnamefont {T.}~\bibnamefont {Rabczuk}},\ }\href
  {\doibase 10.1039/C3NR05991J} {\bibfield  {journal} {\bibinfo  {journal}
  {Nanoscale}\ }\textbf {\bibinfo {volume} {6}},\ \bibinfo {pages} {3618}
  (\bibinfo {year} {2014})}\BibitemShut {NoStop}%
\bibitem [{\citenamefont {Klemens}(1958)}]{Klemens:Book58_Thermal}%
  \BibitemOpen
  \bibfield  {author} {\bibinfo {author} {\bibfnamefont {P.~G.}\ \bibnamefont
  {Klemens}},\ }in\ \href {\doibase 10.1016/S0081-1947(08)60551-2} {\emph
  {\bibinfo {booktitle} {Solid State Physics}}},\ Vol.~\bibinfo {volume} {7},\
  \bibinfo {editor} {edited by\ \bibinfo {editor} {\bibfnamefont
  {F.}~\bibnamefont {Seitz}}\ and\ \bibinfo {editor} {\bibfnamefont
  {D.}~\bibnamefont {Turnbull}}}\ (\bibinfo  {publisher} {Academic Press},\
  \bibinfo {year} {1958})\ pp.\ \bibinfo {pages} {1--98}\BibitemShut {NoStop}%
\bibitem [{\citenamefont {Kresse}\ and\ \citenamefont
  {Furthm\"uller}(1996)}]{Kresse:PRB96_Iterative}%
  \BibitemOpen
  \bibfield  {author} {\bibinfo {author} {\bibfnamefont {G.}~\bibnamefont
  {Kresse}}\ and\ \bibinfo {author} {\bibfnamefont {J.}~\bibnamefont
  {Furthm\"uller}},\ }\href {\doibase 10.1103/PhysRevB.54.11169} {\bibfield
  {journal} {\bibinfo  {journal} {Phys. Rev. B}\ }\textbf {\bibinfo {volume}
  {54}},\ \bibinfo {pages} {11169} (\bibinfo {year} {1996})}\BibitemShut
  {NoStop}%
\bibitem [{\citenamefont {Cullen}\ \emph {et~al.}(2010)\citenamefont {Cullen},
  \citenamefont {Yamamoto}, \citenamefont {Burson}, \citenamefont {Chen},
  \citenamefont {Jang}, \citenamefont {Li}, \citenamefont {Fuhrer},\ and\
  \citenamefont {Williams}}]{Cullen:PRL10_High}%
  \BibitemOpen
  \bibfield  {author} {\bibinfo {author} {\bibfnamefont {W.~G.}\ \bibnamefont
  {Cullen}}, \bibinfo {author} {\bibfnamefont {M.}~\bibnamefont {Yamamoto}},
  \bibinfo {author} {\bibfnamefont {K.~M.}\ \bibnamefont {Burson}}, \bibinfo
  {author} {\bibfnamefont {J.~H.}\ \bibnamefont {Chen}}, \bibinfo {author}
  {\bibfnamefont {C.}~\bibnamefont {Jang}}, \bibinfo {author} {\bibfnamefont
  {L.}~\bibnamefont {Li}}, \bibinfo {author} {\bibfnamefont {M.~S.}\
  \bibnamefont {Fuhrer}}, \ and\ \bibinfo {author} {\bibfnamefont {E.~D.}\
  \bibnamefont {Williams}},\ }\href {\doibase 10.1103/PhysRevLett.105.215504}
  {\bibfield  {journal} {\bibinfo  {journal} {Phys. Rev. Lett.}\ }\textbf
  {\bibinfo {volume} {105}},\ \bibinfo {pages} {215504} (\bibinfo {year}
  {2010})}\BibitemShut {NoStop}%
\bibitem [{\citenamefont {Jiang}\ \emph {et~al.}(2013)\citenamefont {Jiang},
  \citenamefont {Qi}, \citenamefont {Park},\ and\ \citenamefont
  {Rabczuk}}]{Jiang:Nanotech13_Elastic}%
  \BibitemOpen
  \bibfield  {author} {\bibinfo {author} {\bibfnamefont {J.-W.}\ \bibnamefont
  {Jiang}}, \bibinfo {author} {\bibfnamefont {Z.}~\bibnamefont {Qi}}, \bibinfo
  {author} {\bibfnamefont {H.~S.}\ \bibnamefont {Park}}, \ and\ \bibinfo
  {author} {\bibfnamefont {T.}~\bibnamefont {Rabczuk}},\ }\href {\doibase
  10.1088/0957-4484/24/43/435705} {\bibfield  {journal} {\bibinfo  {journal}
  {Nanotechnology}\ }\textbf {\bibinfo {volume} {24}},\ \bibinfo {pages}
  {435705} (\bibinfo {year} {2013})}\BibitemShut {NoStop}%
\bibitem [{\citenamefont {Persson}\ \emph {et~al.}(2010)\citenamefont
  {Persson}, \citenamefont {Lorenz},\ and\ \citenamefont
  {Volokitin}}]{Persson:EPJE10_Heat}%
  \BibitemOpen
  \bibfield  {author} {\bibinfo {author} {\bibfnamefont {B.}~\bibnamefont
  {Persson}}, \bibinfo {author} {\bibfnamefont {B.}~\bibnamefont {Lorenz}}, \
  and\ \bibinfo {author} {\bibfnamefont {A.}~\bibnamefont {Volokitin}},\ }\href
  {\doibase 10.1140/epje/i2010-10543-1} {\bibfield  {journal} {\bibinfo
  {journal} {Eur. Phys. J. E}\ }\textbf {\bibinfo {volume} {31}},\ \bibinfo
  {pages} {3} (\bibinfo {year} {2010})}\BibitemShut {NoStop}%
\bibitem [{\citenamefont {Persson}\ and\ \citenamefont
  {Ueba}(2010)}]{Persson:JPCM10_Heat}%
  \BibitemOpen
  \bibfield  {author} {\bibinfo {author} {\bibfnamefont {B.}~\bibnamefont
  {Persson}}\ and\ \bibinfo {author} {\bibfnamefont {H.}~\bibnamefont {Ueba}},\
  }\href {\doibase 10.1088/0953-8984/22/46/462201} {\bibfield  {journal}
  {\bibinfo  {journal} {J. Phys.: Condens. Matter}\ }\textbf {\bibinfo {volume}
  {22}},\ \bibinfo {pages} {462201} (\bibinfo {year} {2010})}\BibitemShut
  {NoStop}%
\bibitem [{\citenamefont {Ong}\ and\ \citenamefont
  {Pop}(2010)}]{Ong:PRB10_Molecular}%
  \BibitemOpen
  \bibfield  {author} {\bibinfo {author} {\bibfnamefont {Z.-Y.}\ \bibnamefont
  {Ong}}\ and\ \bibinfo {author} {\bibfnamefont {E.}~\bibnamefont {Pop}},\
  }\href {\doibase 10.1103/PhysRevB.81.155408} {\bibfield  {journal} {\bibinfo
  {journal} {Phys. Rev. B}\ }\textbf {\bibinfo {volume} {81}},\ \bibinfo
  {pages} {155408} (\bibinfo {year} {2010})}\BibitemShut {NoStop}%
\bibitem [{\citenamefont {Regner}\ \emph {et~al.}(2013)\citenamefont {Regner},
  \citenamefont {Sellan}, \citenamefont {Su}, \citenamefont {Amon},
  \citenamefont {McGaughey},\ and\ \citenamefont
  {Malen}}]{Regner:NComm13_Broadband}%
  \BibitemOpen
  \bibfield  {author} {\bibinfo {author} {\bibfnamefont {K.~T.}\ \bibnamefont
  {Regner}}, \bibinfo {author} {\bibfnamefont {D.~P.}\ \bibnamefont {Sellan}},
  \bibinfo {author} {\bibfnamefont {Z.}~\bibnamefont {Su}}, \bibinfo {author}
  {\bibfnamefont {C.~H.}\ \bibnamefont {Amon}}, \bibinfo {author}
  {\bibfnamefont {A.~J.~H.}\ \bibnamefont {McGaughey}}, \ and\ \bibinfo
  {author} {\bibfnamefont {J.~A.}\ \bibnamefont {Malen}},\ }\href {\doibase
  10.1038/ncomms2630} {\bibfield  {journal} {\bibinfo  {journal} {Nature
  Commun.}\ }\textbf {\bibinfo {volume} {4}},\ \bibinfo {pages} {1640}
  (\bibinfo {year} {2013})}\BibitemShut {NoStop}%
\bibitem [{\citenamefont {Song}\ \emph {et~al.}(2012)\citenamefont {Song},
  \citenamefont {Reizer},\ and\ \citenamefont {Levitov}}]{Song:PRL12_Disorder}%
  \BibitemOpen
  \bibfield  {author} {\bibinfo {author} {\bibfnamefont {J.~C.~W.}\
  \bibnamefont {Song}}, \bibinfo {author} {\bibfnamefont {M.~Y.}\ \bibnamefont
  {Reizer}}, \ and\ \bibinfo {author} {\bibfnamefont {L.~S.}\ \bibnamefont
  {Levitov}},\ }\href {\doibase 10.1103/PhysRevLett.109.106602} {\bibfield
  {journal} {\bibinfo  {journal} {Phys. Rev. Lett.}\ }\textbf {\bibinfo
  {volume} {109}},\ \bibinfo {pages} {106602} (\bibinfo {year}
  {2012})}\BibitemShut {NoStop}%
\bibitem [{\citenamefont {Seol}\ \emph {et~al.}(2010)\citenamefont {Seol},
  \citenamefont {Jo}, \citenamefont {Moore}, \citenamefont {Lindsay},
  \citenamefont {Aitken}, \citenamefont {Pettes}, \citenamefont {Li},
  \citenamefont {Yao}, \citenamefont {Huang}, \citenamefont {Broido} \emph
  {et~al.}}]{Seol:Science10_Two}%
  \BibitemOpen
  \bibfield  {author} {\bibinfo {author} {\bibfnamefont {J.~H.}\ \bibnamefont
  {Seol}}, \bibinfo {author} {\bibfnamefont {I.}~\bibnamefont {Jo}}, \bibinfo
  {author} {\bibfnamefont {A.~L.}\ \bibnamefont {Moore}}, \bibinfo {author}
  {\bibfnamefont {L.}~\bibnamefont {Lindsay}}, \bibinfo {author} {\bibfnamefont
  {Z.~H.}\ \bibnamefont {Aitken}}, \bibinfo {author} {\bibfnamefont {M.~T.}\
  \bibnamefont {Pettes}}, \bibinfo {author} {\bibfnamefont {X.}~\bibnamefont
  {Li}}, \bibinfo {author} {\bibfnamefont {Z.}~\bibnamefont {Yao}}, \bibinfo
  {author} {\bibfnamefont {R.}~\bibnamefont {Huang}}, \bibinfo {author}
  {\bibfnamefont {D.}~\bibnamefont {Broido}},  \emph {et~al.},\ }\href
  {\doibase 10.1126/science.1184014} {\bibfield  {journal} {\bibinfo  {journal}
  {Science}\ }\textbf {\bibinfo {volume} {328}},\ \bibinfo {pages} {213}
  (\bibinfo {year} {2010})}\BibitemShut {NoStop}%
\end{thebibliography}%

\end{document}